\documentclass[journal]{IEEEtran}
\usepackage{graphicx}
\usepackage{booktabs}
\usepackage{subfigure}
\usepackage{fontawesome}
\usepackage{overpic}
\usepackage{amsmath}
\usepackage{amssymb}
\usepackage{multirow}
\usepackage{makecell}
\usepackage{array}
\usepackage{stfloats}
\usepackage{listings}
\usepackage{algorithm}
\usepackage{color, soul}
\usepackage{bbding}
\usepackage[table]{xcolor}
\usepackage{graphicx}
\usepackage{algpseudocode}

\usepackage{tabularx}

\usepackage{hyperref}
\hypersetup{
	colorlinks=true,
	linkcolor=blue,
	urlcolor=blue,
	citecolor=blue
}

\begin{document}
\title{SS-MAE: Spatial-Spectral Masked Auto-Encoder for Multi-Source Remote Sensing Image Classification}
\author{Junyan Lin,
        Feng Gao, \emph{Member}, \emph{IEEE},
        Xiaochen Shi,
        Junyu Dong, \emph{Member}, \emph{IEEE},\\
        Qian Du, \emph{Fellow}, \emph{IEEE}
\thanks{This work was supported in part by the National Key R\&D Program of China under Grant 2022ZD0117202, and in part by the Natural Science Foundation of Qingdao under Grant 23-2-1-222-zyyd-jch. (Corresponding author: Feng Gao)

Junyan Lin, Feng Gao, Xiaochen Shi, and Junyu Dong are with the School of Computer Science and Technology, Ocean University of China, Qingdao 266100, China. 

Qian Du is with the Department of Electrical and Computer Engineering, Mississippi State University, Starkville, MS 39762 USA.}
}

\markboth{}
{Shell}

\maketitle

\begin{abstract}

Masked image modeling (MIM) is a highly popular and effective self-supervised learning method for image understanding. Existing MIM-based methods mostly focus on spatial feature modeling, neglecting spectral feature modeling. Meanwhile, existing MIM-based methods use Transformer for feature extraction, some local or high-frequency information may get lost. To this end, we propose a spatial-spectral masked auto-encoder (SS-MAE) for HSI and LiDAR/SAR data joint classification. Specifically, SS-MAE consists of a spatial-wise branch and a spectral-wise branch. The spatial-wise branch masks random patches and reconstructs missing pixels, while the spectral-wise branch masks random spectral channels and reconstructs missing channels. Our SS-MAE fully exploits the spatial and spectral representations of the input data. Furthermore, to complement local features in the training stage, we add two lightweight CNNs for feature extraction. Both global and local features are taken into account for feature modeling. To demonstrate the effectiveness of the proposed SS-MAE, we conduct extensive experiments on three publicly available datasets. Extensive experiments on three multi-source datasets verify the superiority of our SS-MAE compared with several state-of-the-art baselines. The source codes are available at \url{https://github.com/summitgao/SS-MAE}.

\end{abstract}


\IEEEpeerreviewmaketitle

\section{Introduction}

\IEEEPARstart{W}ITH the advancement of Earth observation technologies and satellite sensor platforms, significant amounts of remote sensing data have been obtained for 
a variety of purposes, such as vegetation mapping \cite{lihengkai22tgrs}, coastal wetland classification \cite{cui22grsl}, object detection \cite{anomalydetect23tgrs}, change detection \cite{diffchange23tgrs}, land cover classification \cite{10078892}, etc. Among these applications,  land cover classification is fundamental to many operational mapping and reporting natural resource management programs.

\begin{figure}[htbp]
\centering
\includegraphics [width=3.2in]{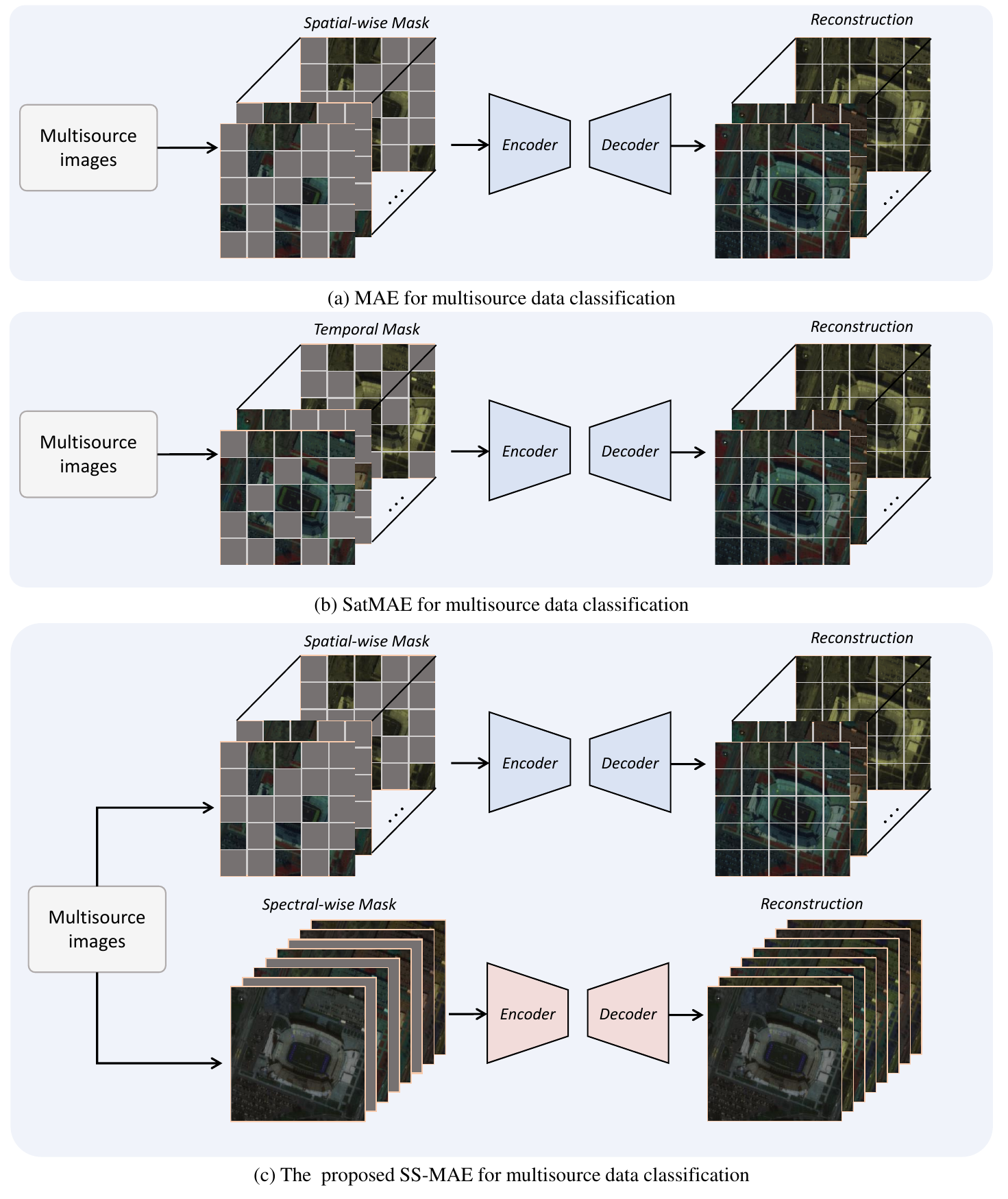}
\caption{Comparisons of the traditional MIMs with the proposed SS-MAE. (a) MAE for multisource data classification. (b) SatMAE for multisource data classification (c) The proposed SS-MAE.}
\label{fig_comp}
\end{figure}

In recent decades, substantial efforts have been made to develop effective classifiers for remote sensing data classification using hyperspectral images (HSIs) \cite{10070826} \cite{10075072} \cite{hg23tgrs} or PolSAR images \cite{9963700} \cite{9930788} \cite{9916296}. Most of these methods use deep neural networks for remote sensing data classification. Multiscale feature fusion \cite{diffchange23tgrs}, dual-branch network \cite{dualview23tgrs}, hypergraph convolution \cite{hypergraph23tgrs} \cite{9940932}, Transformer-like network \cite{10075631}, attention multihop graph \cite{multihopgraph23tgrs}, and neural architecture search \cite{9928571} have been used for hyperspectral or SAR data feature extraction. Although impressive results have been achieved, single modality classification can be problematic in some cases. For example, it is challenging to accurately identify different classes made up of the same material using a hyperspectral image alone (such as roof and cement pavement), as they have similar spectrum corresponds. At the same time, height information from light detection and ranging (LiDAR) data can make up for the limitation. Furthermore, synthetic aperture radar (SAR) sensor possesses the inherent capability of conducting observations around the clock and under any weather conditions. This feature significantly compensates for the shortcomings of hyperspectral sensors during unfavorable weather conditions \cite{6504845}.
Therefore, in this paper, we mainly focus on developing a robust land cover classification method via hyperspectral and LiDAR/SAR remote sensing image fusion.

Many decision-level fusion \cite{liao2014combining} \cite{bigdeli15} \cite{jia21decision} and feature-level fusion \cite{chen2017deep} \cite{xu2017multisource}  \cite{li2020a3clnn} methods have been devoted for multi-source remote sensing image classification, and achieved promising performance. However, these methods commonly employ supervised training strategy, which requires substantial amounts of labeled data, making it a labor-intensive process. Afterwards, unsupervised methods have been developed to alleviate the data annotation problem in multi-source remote sensing image classification \cite{jia23cl} \cite{wang23nncl}. It can directly learn general feature representations from unlabeled data. Recently, masked image modeling (MIM) has demonstrated great potential in unsupervised visual representation learning \cite{bao2021beit} \cite{he2022masked} \cite{Girdhar_2023_CVPR}. It learns universal feature representations from unlabeled data by reconstructing masked image patches, which can be used to achieve powerful performance in downstream tasks. However, MIM has rarely been considered in multi-source remote sensing image classification, and in this paper, we focus on developing a robust multi-source image classification method based on MIM.

There are two challenges associated with the development of MIM-based multi-source remote sensing image classification methods: \textbf{1) Spectral feature modeling.} As shown in Fig. \ref{fig_comp}, the existing MIM techniques that can be directly employed for multi-source remote sensing image classification can be divided into two categories. The first category is spatial-wise masking employed by MAE \cite{he2022masked}, as illustrated in Fig. \ref{fig_comp}(a), which masks random patches in the spatial domain, and ignores the spectral information. This scheme does not fully explore the spectral information, leading to limited classification performance improvement. The second category is temporal masking employed by SatMAE \cite{cong2022satmae}, which takes spectral information into account, as illustrated in Fig. \ref{fig_comp}(b). However, the way of masking random patches band by band may lead to information leakage, thus reducing the difficulty of reconstruction. Therefore, how to effectively integrate spectral features for MIM-based methods poses the main challenge for us. \textbf{2) Local and global feature extraction. } Transformer is commonly used for MIM, and it tends to pay more attention on global feature dependencies via long-range interactions. However, some local features or high-frequency information (such as edge and texture) may get lost. Consequently, how to simultaneously modeling the local and global features in transformer-like network is a tough challenge.

To tackle these challenges, we present a  \textit{Spatial-Spectral Masked Auto-Encoder} (SS-MAE for short) for HSI and LiDAR/SAR data joint classification. Specifically, we adopt a new MIM technique as illustrated in Fig. \ref{fig_comp}, which comprises spatial-wise branch and spectral-wise branch. The spatial-wise branch masks random patches and reconstructs missing pixels, while the spectral-wise branch masks random spectral channels and reconstructs missing channels. Hence, the proposed SS-MAE fully exploits the spatial and spectral feature representations of the input multi-source data. Furthermore, to complement local features in the training stage, we add two lightweight CNNs for feature extraction. Both global and local features are used in feature modeling. To demonstrate the effectiveness of the proposed SS-MAE, we conduct extensive experiments on three publicly available datasets. The result shows that the proposed SS-MAE outperforms several state-of-the-art methods.

Our main contributions can be summarized as follows:

\begin{itemize}

\item We propose SS-MAE, which performs both spatial and spectral reconstructions for feature representation learning. To the best of our knowledge, this is the first MIM work for multi-source remote sensing image classification.
    
\item We propose a Transformer and CNN hybrid network for global and local feature modeling. Dual-attention Transformer is employed for both spectral and spatial feature pretraining. Moreover, lightweight CNNs are used to capture local features.

\item Extensive experiments on three benchmark datasets reveal that our proposed method outperforms the state-of-the-art methods. As a byproduct, we will release the codes of SS-MAE to benefit other researchers. 

\end{itemize}

The remainder of this paper is organized as follows: In Section II, we review closely related multi-source remote sensing image classification methods and self-supervised methods. The details of SS-MAE are described in Section III. Experiments on three multi-source datasets are presented in Section IV. Conclusions are drawn in Section V.

\section{Related Works}

\subsection{Supervised Multi-Source Remote Sensing 
 Data Classification} 

There are two types of supervised multi-source classification methods: decision-fusion \cite{liao2014combining} \cite{bigdeli15} \cite{jia21decision} and feature-fusion \cite{chen2017deep} \cite{xu2017multisource}  \cite{li2020a3clnn}. Specifically, decision-fusion methods usually use several classifiers separately and then combine the classifications together to form the final classification result. Liao et al. \cite{liao2014combining} adopted four support vector machines (SVM) as the classifier for spectral, spatial, elevation, and graph fused features and combine the classification results by weighted summation. The combination weights are determined by the classification accuracy of each source. Ge et al. \cite{ge2019hyperspectral} gained extinction profile and local binary pattern features from both sources and classified them by collaborative representation-based classifier with Tikhonov regularization. With the popularity of deep learning, feature-fusion methods have been used more frequently. Most feature-fusion methods employ dual-stream networks to extract features from multi-source remote sensing images and then converge the extracted features through various interaction approaches. Li et al. \cite{li2020a3clnn} presented a dual-channel framework, amalgamating spatial, spectral, and multiscale attention mechanisms. This model was devised to extract features and classify multisource remote sensing data. Gao et al. \cite{gaoyh22tgrs} designed a depthwise cross-attention module to extract not only self-correlation but also cross-correlation from diverse multisource data. Wang et al. \cite{9698196} proposed AM$^3$Net which includes an involution operator, a spectral-spatial mutual-guided module, and a spectral-spatial mutual-guided module.

Although these supervised learning-based methods have achieved excellent performance, their performance is highly dependent on the large-scale labeled training samples. Correspondingly, self-supervised learning methods that do not require many training samples may overcome this limitation.

\begin{figure*}[t]
\centering
\includegraphics [width=7in]{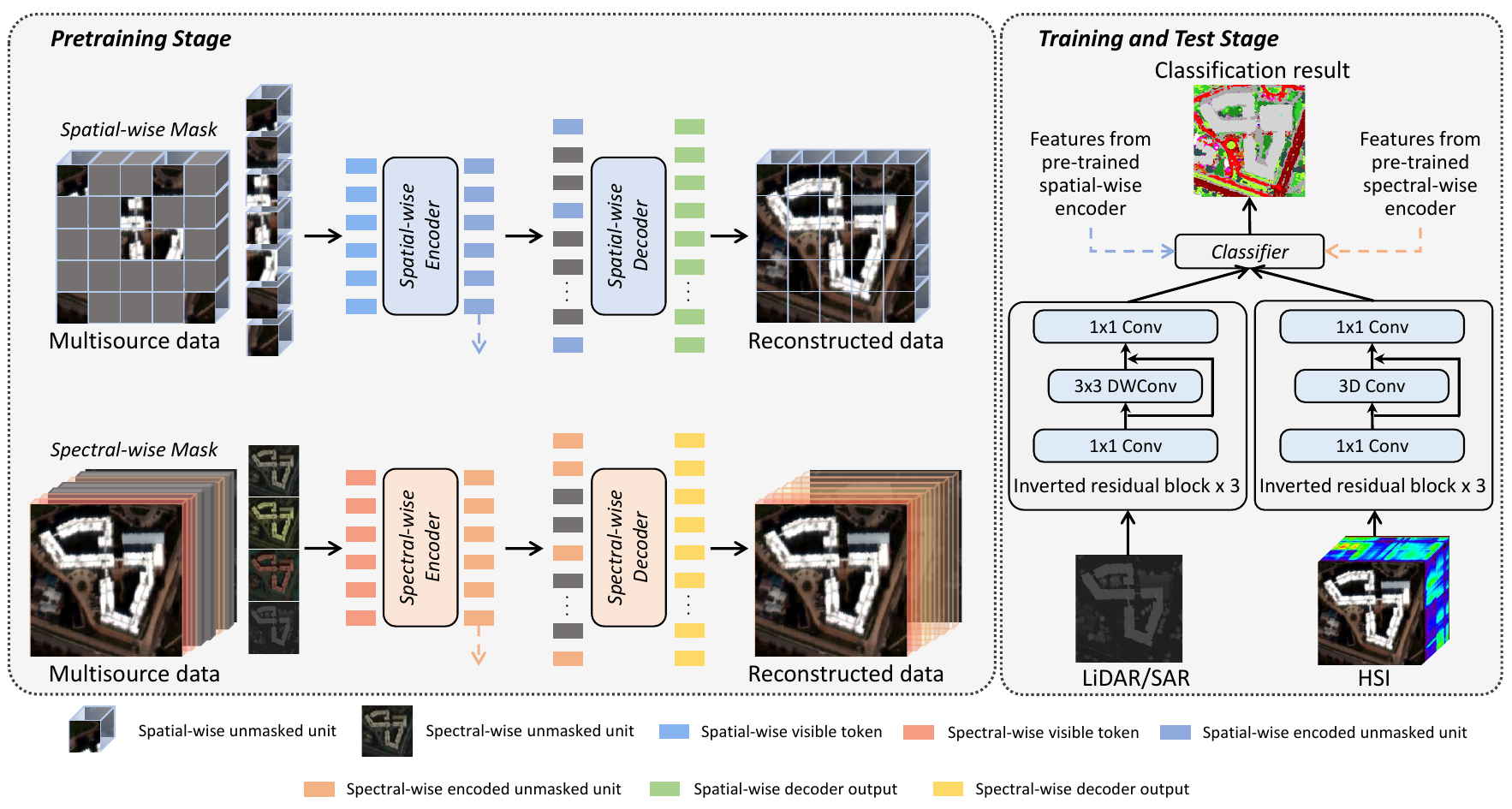}
\caption{Framework of the proposed spatial-spectral masked auto-encoder for multi-source image classification. It is comprised of the pre-training stage, the training and test stage. In the pretraining stage, the network contains a spatial-wise branch and a spectral-wise branch. The spatial-wise branch masks random patches and reconstructs missing pixels, while the spectral-wise branch masks random spectral channels and reconstructs missing channels. In the training and test stage, two lightweight CNNs are built to complement local features. Each lightweight CNN contains three inverted residual blocks, and therefore both global and local features are used in feature representation.}
\label{framework}
\end{figure*}

\subsection{Self-Supervised Learning}

Self-supervised learning can process unlabeled data to obtain robust representations that can facilitate downstream learning tasks. Existing self-supervised learning methods can be divided into contrastive-based \cite{he2020momentum}\cite{chen2020simple}\cite{caron2021emerging}\cite{liu2022multi} and reconstruction-based \cite{bao2021beit}\cite{he2022masked}\cite{baevski2022data2vec}\cite{cong2022satmae}. 

Contrastive-based methods learn feature representations by minimizing the distance between positive samples while maximizing the distance between negative samples. SimCLR \cite{chen2020simple} harnesses the power of contrastive loss to learn visual representations. It achieves this by maximizing agreement between differently augmented views of the identical data sample. Liu et al. \cite{liu2022multi} demonstrated the effectiveness of using contrast learning in remote sensing data classification. Also, Contrastive-based methods are beginning to be used in the multi-modal remote sensing image classification. Feng et al. \cite{10187150} use joint intra- and cross-modal contrastive learning to mine multi-modal feature representations during pre-training. They design a simple yet effective hybrid cross-modal fusion module in the fine-tuning stage to better compactly integrate complementary information across these modalities for ground object classification.

Reconstruction-based methods learn to predict plausible content for the masked regions based on the information provided by the visible regions. The training is commonly guided by a loss function that measures the difference between the predicted completion and the ground truth content in the masked regions. He et al. \cite{he2022masked} proposed a simple yet effective method to pretrain large vision models by masking patches of an image and using an autoencoder to predict the masked patches. Cong et al. \cite{cong2022satmae} presented a reconstruction-based pre-training framework for multi-spectral satellite imagery, called SatMAE. Although yielding excellent performance, the length of token sequences in SatMAE becomes excessively long when dealing with high-dimensional spectral data. This could potentially lead to the challenge of consuming excessive computational resources. Furthermore, the proposed SS-MAE is based on a Transformer and CNN hybrid network, which results in enhanced computational efficiency compared with SatMAE.

To this end, we presented a two-branch auto-encoder framework for spatial and spectral feature modeling, respectively. Considering that the model is reconstructing images that are masked along an entire spatial or spectral dimension, such a framework is preferable for HSI feature modeling. To the best of our knowledge, we are the first to use MIM for multi-source remote sensing classification tasks.

\subsection{Vision Transformers}

Vaswani et al. \cite{vaswani2017attention} introduced the Transformer model, which relies exclusively on attention mechanisms, and it has demonstrated exceptional performance in the field of natural language processing. Recently, Transofrmer-based architecture quickly becomes popular in computer vision \cite{dosovitskiy2020image} \cite{10124835} \cite{10091778} \cite{10091771} and remote sensing \cite{10075631} \cite{9989404} \cite{9874815} \cite{10138021}. In multi-modal remote sensing image classification, transformer-like architectures have been explored recently. Zhang et al. \cite{10145469} present a multimodal transformer network (MTNet) that captures both the specific and shared features of HSI and LiDAR data. Roy et al. \cite{10153685} introduce a multimodal fusion transformer network, which uses cross-patch attention for remote sensing data classification. The cross-patch attention employs complementary information from LiDAR in the transformer encoder to achieve better generalization.

Motivated by the success of Vision Transformers, most of the recent MIM methods utilize the Transformer for feature extraction. Transformer-based methods focus on learning long-range dependencies to obtain better feature representations, and they ignore some local or high-frequency information (such as textures and corners) \cite{park2022vision}. Due to the complexity of ground objects in remote sensing images, local features are essential for discriminative representation. To this end, we add two lightweight CNN in the training stage for local feature extraction.

\begin{algorithm}[b]
\caption{Pseudocode of SS-MAE in a PyTorch-like style. (pre-training stage)}
\label{alg1}

\definecolor{col}{rgb}{0.3,0.75,0.3}
\lstset{
  backgroundcolor=\color{white},
  basicstyle=\fontsize{7.2pt}{7.2pt}\ttfamily\selectfont,
  columns=fullflexible,
  breaklines=true,
  captionpos=b,
  commentstyle=\fontsize{7.2pt}{7.2pt}\color{col},
  keywordstyle=\fontsize{7.2pt}{7.2pt},
}

\begin{lstlisting}[language=python]
# Spa_E, Spe_E: Spatial-wise and Spectral-wise encoders
# Spa_D, Spe_D: Spatial-wise and Spectral-wise decoders
# x1: SAR or LiDAR
# x2: HSI

# load a mini-batch x1 and x2 with N samples
# x1: N x c1 x P x P
# x2: N x c2 x P x P
for x1, x2 in loader:
    # T: N x (c1 + c2) x P x P
    T = concat([x1, x2], dim = 1) 
    TM_spa = spa_mask(T)  # spacial-wise mask
    TM_spe = spe_mask(T)  # spectral-wise mask
    
    #f_spa: N x (P x P) x D, f_spe: N x (c1 + c2) x D
    f_spa = Spa_E.forward(TM_spa.view(N, P * P, c1 + c2))
    f_spe = Spe_E.forward(TM_spe.view(N, c1 + c2, P * P))

    # tr: linear projection and resize
    # R_spa: N x (c1 + c2) x P x P
    R_spa = tr(Spa_D.forward(f_spa)) 
    # R_spe: N x (c1 + c2) x P x P
    R_spe = tr(Spe_D.forward(f_spe)) 

    # MSE loss computation
    loss = MSELoss(R_spa, T) + MSELoss(R_spe, T)

    # back propagation
    loss.backward()
    update(Spa_E.param, Spa_D.param, \
        Spe_E.param, Spe_D.param)
\end{lstlisting}
\end{algorithm}

\section{Methodology} 
\begin{figure}[t]
\centering
\includegraphics [width=3.3in]{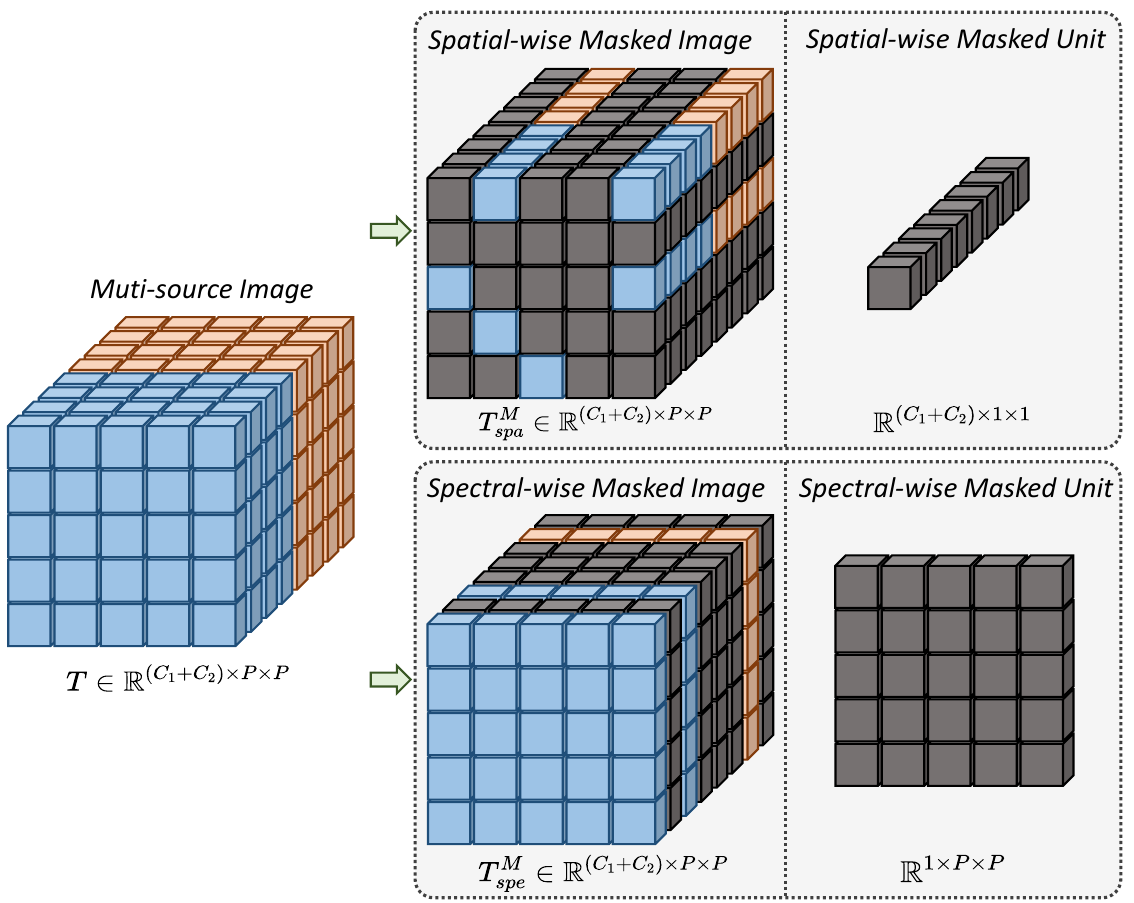}
\caption{The masking strategy of spatial-wise masking and spectral-wise masking.}
\label{masking_strategy}
\end{figure}

In this section, we first describe the framework of the proposed SS-MAE, and then detail the pre-training stage, dual-attention Transformer network, the training and inference stage.

\subsection{Proposed SS-MAE Framework}

The proposed SS-MAE framework is shown in Fig. \ref{framework}, which consists of two parts: the pre-training stage and the training stage. For a given pair of hyperspectral image patch $X_1 \in \mathbb{R}^{C_1 \times P \times P}$ and SAR/LiDAR patch $X_2 \in \mathbb{R}^{C_2 \times P \times P}$ covering the same region, they are first concatenated to $T \in \mathbb{R}^{(C_1+C_2) \times P \times P}$, and then two different masking strategies are applied to $T$, i.e., spatial-wise masking and spectral-wise masking, to generate two different masked images, $T^M_{spa}$, and $T^M_{spe}$. Afterwards, $T^M_{spa}$ and $T^M_{spe}$ are fed into the SS-MAE encoder-decoder to obtain the reconstructed images $R_{spa}$ and $R_{spe}$, respectively. In the training stage, in addition to extracting the global representations of the two-branch MAE encoder, the two Inverted Residual Blocks (IRB) are used to extract the high-frequency information-rich representations of $X_1$ and $X_2$, respectively, and the final classification results are obtained by concatenating these representations with the classifier.

\subsection{Pre-training Stage}

The goal of pre-training is to enable the model to extract feature representations from a mass of unlabeled data, and MAE confirms that mask-reconstruction can easily achieve this. In this paper, we pre-train the model using the similar mask-reconstruction as MAE. Algorithm \ref{alg1} provides the pseudo-code of SS-MAE in pre-training stage.

\textbf{Masking Strategy.} MIM is one of the most prominent pretext tasks in self-supervised learning methods. During pretraining, MIM is implemented by masking a portion of the image and asking the model to reconstruct the image representation or feature representation being masked, thus forcing the model to learn universal features in unlabeled data. Random masking is an essential component of the MIM technique. However, existing methods are often limited to masking along the spatial dimension, ignoring the spectral information. To this end, we propose spatial-spectral masking method for hyperspectral feature representation.

\emph{Spatial-wise masking:} We randomly mask some pixels in the spatial dimension as shown in Fig. \ref{masking_strategy}. For the input $T\in \mathbb{R}^{(C_1+C_2) \times P \times P}$,  the size of each masked unit is $(C_1+C_2) \times 1 \times 1 $. Hence, we obtain the spatial-wise masked image $T^M_{spa}$.

\emph{Spectral-wise masking:} As the responses of different spectral bands for the same object are sometimes similar, the challenge of reconstruction is reduced if different spectra are spatially complementary to each other. As such, we mask the whole band data for spectral-wise masking. Specifically, for the input $T\in \mathbb{R}^{(C_1+C_2) \times P \times P}$, we compute a spectral-wise masked image $T^M_{spe}$ by randomly masking some spectral bands. The size of each masked unit is $1 \times P \times P $, as shown in Fig. \ref{masking_strategy}. The masking ratio is a critical parameter that may affect the classification performance. Therefore, we conducted comprehensive experiments in Section IV. B to select the best masking ratio. 

It should be noted that spectral-wise masking is not applicable to single-band data such as LiDAR. Consequently, for HSI and LiDAR data classification, we deliberately exclude LiDAR in spectral-wise masking. However, for HSI and multispectral data joint classification, or HSI and PolSAR data joint classification, spectral-wise masking is essential for the proposed SS-MAE.

\textbf{Reconstruction Target.}  Our SS-MAE reconstructs the input by predicting the pixel values in spatial-wise and spectral-wise masked units. The loss function consists of a spatial reconstruction loss $\mathcal{L}(T, R_{spa})$, and a spectral reconstruction loss $\mathcal{L}(T, R_{spe})$. The whole loss $\mathcal{L}_\textrm{SSMAE}$ is a weighted sum: 
\begin{equation}
\mathcal{L}_\textrm{SSMAE} = \lambda \mathcal{L}(T, R_{spa}) + \mathcal{L}(T, R_{spe}).
\end{equation}
We use mean squared error (MSE) loss for $\mathcal{L}(T, R_{spa})$ and $\mathcal{L}(T, R_{spa})$. $\lambda$ is set to 2 in our experiments. In order to maintain consistency with the loss function of MAE, we only compute the loss pertaining to masked pixels, while disregarding those that are unmasked.

\begin{algorithm}[t]
\caption{Pseudocode of SS-MAE in a PyTorch-like style. (training stage)}
\label{alg2}

\definecolor{col}{rgb}{0.3,0.75,0.3}
\lstset{
  backgroundcolor=\color{white},
  basicstyle=\fontsize{7.2pt}{7.2pt}\ttfamily\selectfont,
  columns=fullflexible,
  breaklines=true,
  captionpos=b,
  commentstyle=\fontsize{7.2pt}{7.2pt}\color{col},
  keywordstyle=\fontsize{7.2pt}{7.2pt},
}

\begin{lstlisting}[language=python]
# Spa_E, Spe_E: Spatial-wise and Spectral-wise encoders
# x1: SAR or LiDAR
# x2: HSI
# labels: labels in the dataset
# IRB: Inverted residual block

# load a mini-batch x1 and x2 with N samples
# x1: N x c1 x P x P
# x2: N x c2 x P x P
for x1, x2 in loader:
    # T: N x (c1 + c2) x P x P
    T = concat([x1, x2], dim = 1) 

    # f_spa: N x (PxP) x D, f_spe: N x (c1 + c2) x D
    f_spa = Spa_E.forward(T.view(N, P * P, c1 + c2))
    f_spe = Spe_E.forward(T.view(N, c1 + c2, P * P))
    
    # tr: linear projection and resize
    I_1 = tr(IRB_1.forward(x1)) # I_1: N x 256 x D
    I_2 = tr(IRB_2.forward(x2)) # I_2: N x 256 x D

    # fn: N x (c1 + c2 + PxP + 512) x D
    fn = concat(f_spa + f_spe + I_1 + I_2)

    # r: N x Cnum, Cnum is the number of categories
    r = classifier(fn)
    
    # CrossEntropy loss computation
    loss = CrossEntropyLoss(r, labels) # labels: N x Cnum

    # back propagation
    loss.backward()
    update(Spa_E.param, Spe_E.param, \
    IRB_1.param, IRB_2.param, classifier.param)
\end{lstlisting}
\end{algorithm}

\begin{figure*}[htb]
    \centering
    \includegraphics[width=0.9\textwidth]{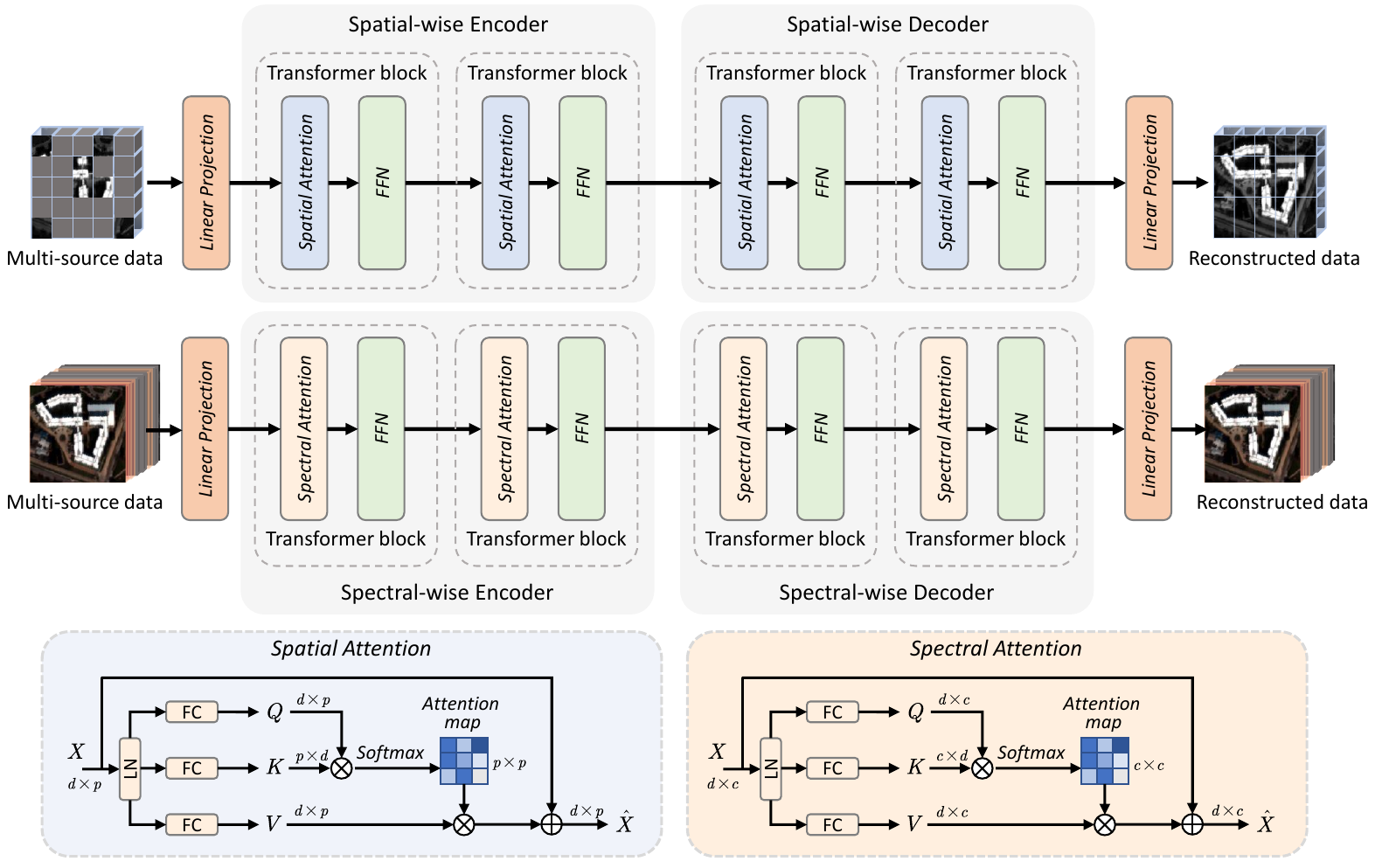}
    \caption{Architecture of the dual-attention Transformer. Spatial-wise attention and spectral-wise attention are considered for masked information reconstruction. Therefore, the model fully exploits the spatial and spectral representations of the input multi-source data.}
    \label{fig:former}
\end{figure*}

\subsection{Dual-Attention Transformer}

Leveraging the natural property that both masked and unmasked units can serve as input tokens, we propose a dual-attention Transformer encoder-decoder architecture to accomplish the task of reconstructing $T^M_{spa}$ and $T^M_{spe}$ into $T$, as shown in Fig. \ref{fig:former}. 

\textbf{Spatial-Wise Reconstruction.} Only visible tokens in $T^M_{spa}$ are retained, and these tokens are handled by linear projection to generate $T^{M'}_{spa} \in \mathbb{R}^{d\times p}$. Here $p$ is the number of visible tokens, and $d$ is the feature dimension of each token. Next, $T^{M'}_{spa}$ is fed into the spatial-wise encoder with $\mathcal{B}$ Transformer blocks to compute the encoded visible tokens.

After that, full image tokens (consisting of both encoded visible tokens and mask tokens) are fed into the spatial-wise decoder to accomplish the reconstruction task. Positional embeddings are used to encode the token locations in the image. The decoder has another $\mathcal{B}$ Transformer blocks. 

Each Transformer block is composed of spatial-wise attention computation and feed-forward neural network (FFN). For spatial-wise attention, given $X\in \mathbb{R}^{d\times p}$ from the input, we first apply $1\times1$ convolutions to enhance the input feature, and obtain query $Q$, key $K$, and value $V$. After that, $Q$ and $K$ are reshaped to conduct dot-product to generate spatial-wise attention map $A \in \mathbb{R}^{p\times p}$. Overall, the spatial attention is defined as:
\begin{equation}
  \hat{X}=\text{Attention}\left(\hat{Q}, \hat{K}, \hat{V}\right) + X,
\end{equation}
\begin{equation}
  \text{Attention}\left(\hat{Q}, \hat{K}, \hat{V}\right)= \hat{V} \text{Softmax}\left(\hat{K}\hat{Q}/\alpha\right),
\end{equation}
where $X$ and $\hat{X}$ are the input and output feature maps. $\hat{Q}$, $\hat{K}$, and $\hat{V}$ are reshaped tensors from the input. $\alpha$ is a learnable scaling parameter to control the softmax. To achieve multi-head attention, we split the $Q$, $K$, and $V$ into $\mathcal{H}$ heads along the feature channel dimension, respectively. The dimension of each head is $p/\mathcal{H}$, and hence separate attention maps can be learned efficiently in parallel. We adopt the identical FFN architecture as proposed in ViT \cite{dosovitskiy2020image}. This FFN consists of two linear layers, with a GELU activation function separating them. The first layer expands the dimension by a factor of 4, while the subsequent layer reduces it by the same ratio.

\textbf{Spectral-Wise Reconstruction.} Here visible tokens are spectral bands, and the mask tokens are masked spectral bands. Similarly, only visible tokens are retained and fed into the encoder to compute the encoded visible tokens. In spectral-wise decoder, both the visible tokens and mask tokens are fed into the decoder for the reconstruction task. 

It should be noted that after linear projection, $T^M_{spa}$ is transformed to $T^{M'}_{spe}\in\mathbb{R}^{d\times c}$. Here $c$ is the number of visible spectral bands, and $d$ is the feature dimension of each band after linear projection. In spatial attention, the attention map computes the similarity of visible pixels, while  in spectral attention, the attention map computes the similarity of visible spectral bands. 

\subsection{Training and Optimization}

Conventional masked autoencoders only fine-tune the encoder parameters on the training data. Although it is convenient and easy to implement, the model may perform sub-optimally for hyperspectral data. Specifically, Transformer tends to pay more attention on long-range dependencies of the features and the semantics of the input data. Some local features or high-frequency information (such as edge and texture) may get lost. 

In order to add local features, we add two lightweight CNN to capture local features. For LiDAR/SAR data, the network consists of three inverted residual blocks (IRB). As shown in Fig. \ref{framework}, the IRB is composed of a projection layer, a depth-wise convolution layer, and an expansion projection layer.  It is defined as follows:

\begin{equation}
    \text{IRB}(X) = W_1(\mathcal{F}(W_2(X))),
\end{equation}
\begin{equation}
    \mathcal{F}(X)=\text{DWConv}(X)+X,
\end{equation}
where $W_1$ and $W_2$ are $1\times1$ convolution, $\text{DWConv}(\cdot)$ is $3\times3$ depth-wise convolution. We include the GELU activation  and batch normalization after the first and second convolution layers. The depth-wise convolution is capable of extracting local features with negligible extra computational cost. 

For HSI feature extraction, the IRB is composed of a projection layer, a 3D convolution layer, and an expansion projection layer. The difference from IRB for LiDAR/SAR data is the replacement of the depth-wise convolution with a 3D convolution. The expansion ratio is 2 in our implementations. 

Finally, the features from spatial-wise and spectral-wise Transformer encoders are stacked with features from IRBs, as shown in Fig. \ref{framework}. Both global and local information from the input multisource data are integrated for classification.  Algorithm \ref{alg2} shows the pseudocode of SS-MAE for the training stage.

\section{Experimental Results and Analysis}

\subsection{Dataset Description}

To evaluate the effectiveness of our proposed SS-MAE, we employ it on three multi-source remote sensing datasets: Berlin, Augsburg, and Houston 2018. Berlin and Augsburg datasets are used for hyperspectral and SAR data classification. Houston 2018 is used for hyperspectral and LiDAR data classification. 

\definecolor{m1}{HTML}{1AA319}  
\definecolor{m2}{HTML}{D8D8D8}  
\definecolor{m3}{HTML}{D85959}  
\definecolor{m4}{HTML}{00CC33}  
\definecolor{m5}{HTML}{CC9934}  
\definecolor{m6}{HTML}{F4E701}  
\definecolor{m7}{HTML}{CC66CC}  
\definecolor{m8}{HTML}{0035FF}  

\begin{table}[htb]
\centering
\caption{The number of training and test samples for the Berlin dataset.}
\begin{tabular}{cccc|cc}
\toprule
    No. & Name & Color & ~ & Train & Test \\
\midrule
    1 & Forest & \cellcolor{m1} & & 443 & 54511\\
    2 & Residential area & \cellcolor{m2} & & 423 & 268219 \\
    3 & Industrial area & \cellcolor{m3} & & 499 & 19067\\
    4 & Low plants & \cellcolor{m4} & & 376 & 58906 \\
    5 & Soil & \cellcolor{m5} & & 331 & 17095 \\
    6 & Allotment & \cellcolor{m6} & & 280 & 13025\\
    7 & Commercial area & \cellcolor{m7} & & 298 & 24526\\
    8 & Water & \cellcolor{m8} & & 170 & 6502\\
\midrule
    \multicolumn{3}{c}{Total} & & 2820 & 461851\\
\bottomrule
\end{tabular}
 \label{table_berlin}
\end{table}

\definecolor{mm1}{HTML}{1AA319}  
\definecolor{mm2}{HTML}{D8D8D8}  
\definecolor{mm3}{HTML}{D85959}  
\definecolor{mm4}{HTML}{00CC33}  
\definecolor{mm5}{HTML}{F4E701}  
\definecolor{mm6}{HTML}{CC66CC}  
\definecolor{mm7}{HTML}{0035FF}  

\begin{table}[htb]
\centering
\caption{The number of training and test samples for the Augsburg dataset.}
\begin{tabular}{cccc|cc}
\toprule
    No. & Name & Color & ~ & Train & Test \\
\midrule
    1 & Forest & \cellcolor{mm1} & & 146 & 13361\\
    2 & Residential area & \cellcolor{mm2} & & 264 & 30065 \\
    3 & Industrial area & \cellcolor{mm3} & & 21 & 3830\\
    4 & Low plants & \cellcolor{mm4} & & 248 & 26609 \\
    5 & Allotment & \cellcolor{mm5} & & 52 & 523\\
    6 & Commercial area & \cellcolor{mm6} & & 7 & 1638\\
    7 & Water & \cellcolor{mm7} & & 23 & 1507\\
\midrule
    \multicolumn{3}{c}{Total} & & 761 & 77533\\
\bottomrule
\end{tabular}
 \label{table_augsburg}
\end{table}

\definecolor{hh1}{HTML}{32CD33}  
\definecolor{hh2}{HTML}{ADFF30}  
\definecolor{hh3}{HTML}{008081}  
\definecolor{hh4}{HTML}{228B22}  
\definecolor{hh5}{HTML}{2E4F4E}  
\definecolor{hh6}{HTML}{8B4512}  
\definecolor{hh7}{HTML}{00FFFF}  
\definecolor{hh8}{HTML}{FFFFFF}  
\definecolor{hh9}{HTML}{D3D3D3}  
\definecolor{hh10}{HTML}{FE0000}  
\definecolor{hh11}{HTML}{A9A9A9}  
\definecolor{hh12}{HTML}{696969}  
\definecolor{hh13}{HTML}{8B0001}  
\definecolor{hh14}{HTML}{C86400}  
\definecolor{hh15}{HTML}{FEA500}  
\definecolor{hh16}{HTML}{FFFF00}  
\definecolor{hh17}{HTML}{DAA521}  
\definecolor{hh18}{HTML}{FF00FE}  
\definecolor{hh19}{HTML}{0000FE}  
\definecolor{hh20}{HTML}{3FE0D0}  

\begin{table}[tp]
\centering
\caption{The number of training and testing samples for the Houston 2018 dataset.}
\begin{tabular}{cccc|cc} \hline
\toprule
    No. & Name & Color & ~ & Train & Test \\
\midrule
    1 & Health grass & \cellcolor{hh1} & & 1000 & 39196\\
\midrule
    2 & Stressed grass & \cellcolor{hh2} & & 1000 & 130008\\
\midrule
    3 & Artificial turf & \cellcolor{hh3} & & 1000 & 2736\\
\midrule
    4 & Evergreen trees & \cellcolor{hh4} & & 1000 & 54322\\
\midrule
    5 & Deciduous trees & \cellcolor{hh5} & & 1000 & 20172\\
\midrule
    6 & Bare earth & \cellcolor{hh6} & & 1000 & 18064\\
\midrule
    7 & Water & \cellcolor{hh7} & & 500 & 1064\\
\midrule
    8 & Residential buildings & \cellcolor{hh8} & & 1000 & 15899\\
\midrule
    9 & Non-residential buildings & \cellcolor{hh9} & & 1000 & 894769\\
\midrule
    10 & Roads & \cellcolor{hh10} & & 1000 & 183283\\
\midrule
    11 & Sidewalks & \cellcolor{hh11} & & 1000 & 136035\\
\midrule
    12 & Crosswalks & \cellcolor{hh12} & & 1000 & 6059\\
\midrule
    13 & Major thoroughfares & \cellcolor{hh13} & & 1000 & 185438\\
\midrule
    14 & Highways & \cellcolor{hh14} & & 1000 & 39438\\
\midrule
    15 & Railways & \cellcolor{hh15} & & 1000 & 27748\\
\midrule
    16 & Paved parking lots & \cellcolor{hh16} & & 1000 & 45932\\
\midrule
    17 & Unpaved parking lots & \cellcolor{hh17} & & 250 & 587\\
\midrule
    18 & Cars & \cellcolor{hh18} & & 1000 & 26289\\
\midrule
    19 & Trains & \cellcolor{hh19} & & 1000 & 21479\\
\midrule
    20 & Stadium seats & \cellcolor{hh20} & & 1000 & 27296\\
\midrule
    \multicolumn{4}{c|}{Total} & 18750 & 2018910\\
\bottomrule
\end{tabular}
 \label{table_Houston2018}
\end{table}

\subsubsection{Berlin Dataset}
It covers the urban and rural areas of Berlin. The hyperspectral data is simulated EnMAP data based on HyMap HS data. The SAR data is Sentinel-1 dual-Pol (VV–VH) single look complex (SLC) product obtained from the European Space Agency (ESA) \cite{hong2021multimodal}. The HSI has 797 × 220 pixels, 244 spectral bands in the wavelength range of 400–2500 nm \cite{okujeni2016berlin}. The SAR image has 1723 × 476 pixels. The nearest neighbor interpolation was employed to match the spatial resolution of both images.

\subsubsection{Augsburg Dataset}
This dataset is composed of HSI and SAR data from Augsburg, Germany. The HSI was acquired by the HySpex sensor, and the SAR data was acquired by the Sentinel-1 sensor. All images are with the ground sample distance (GSD) of 30 m. The spatial size of both images is 332 × 485 pixels. The HSI has 80 spectral bands ranging from 0.4 to 2.5 $\mu$m, and the SAR data has four features derived from polarization decomposition (VV intensity, VH intensity, real part and imaginary part of the off-diagonal element of the PolSAR covariance matrix).

\begin{figure*}[ht]
\centering
\includegraphics [width=0.99\textwidth]{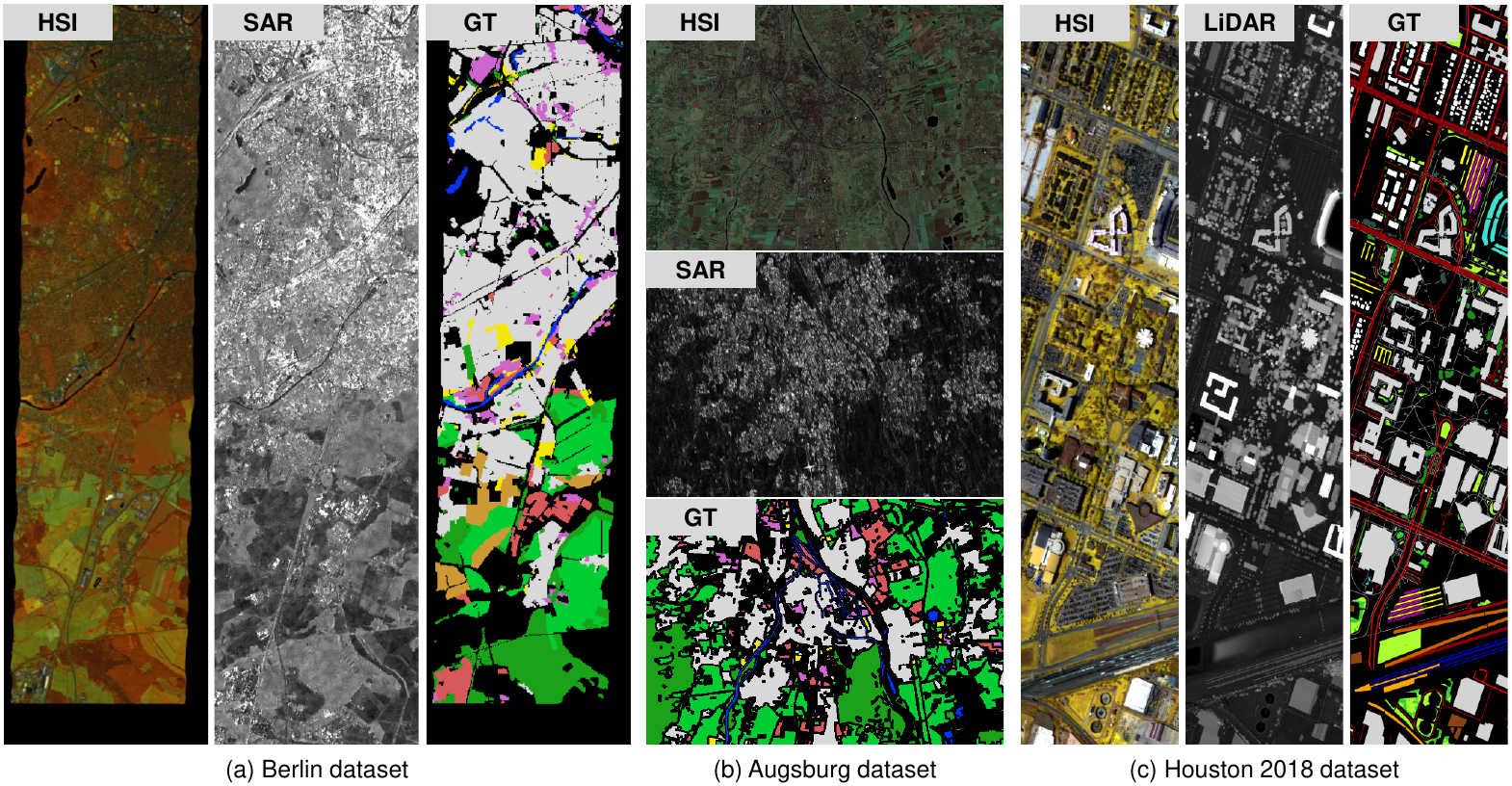}
\caption{Visualization of the datasets. (a) Berlin dataset. (b) Augsburg dataset. (c) Houston 2018 dataset.}
\label{dataset}
\end{figure*}

\begin{figure*}[ht]
\centering
\subfigure[]{\includegraphics[width=.32\textwidth]{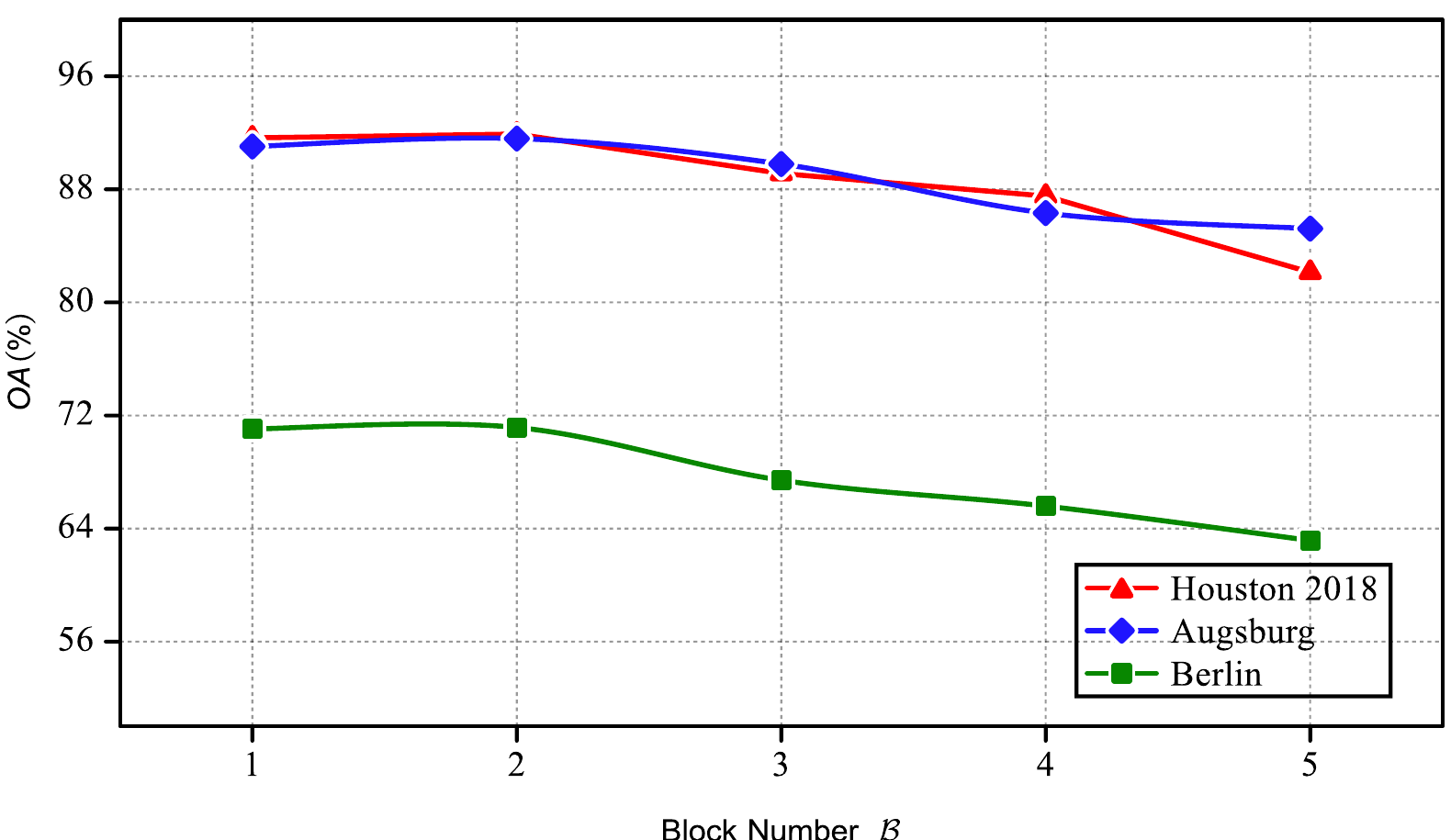}}
\subfigure[]{\includegraphics[width=.32\textwidth]{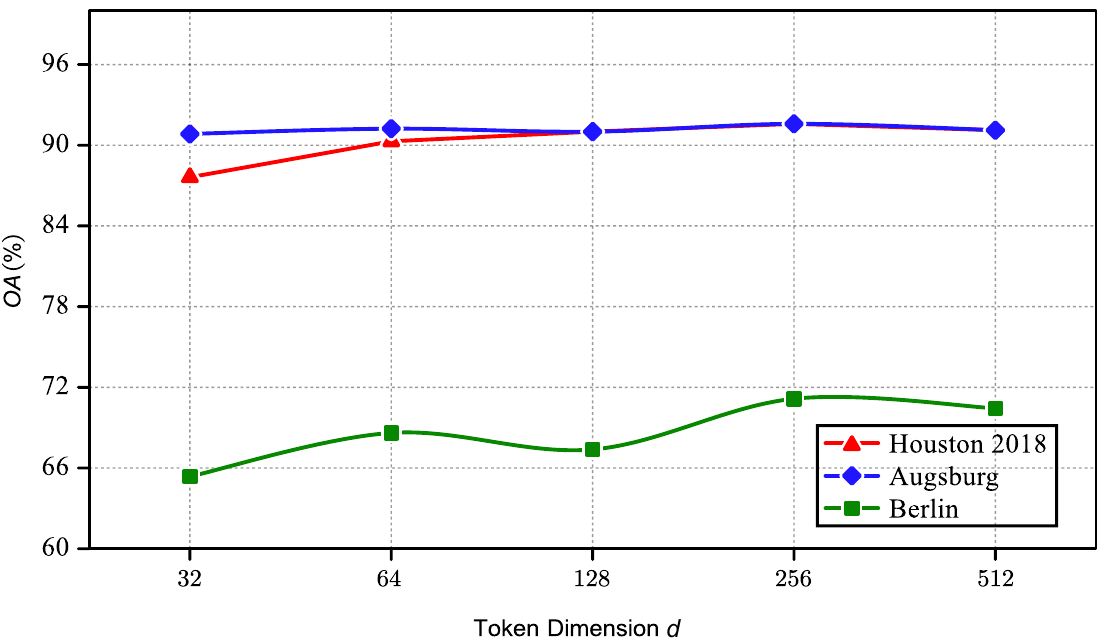}}
\subfigure[]{\includegraphics[width=.32\textwidth]{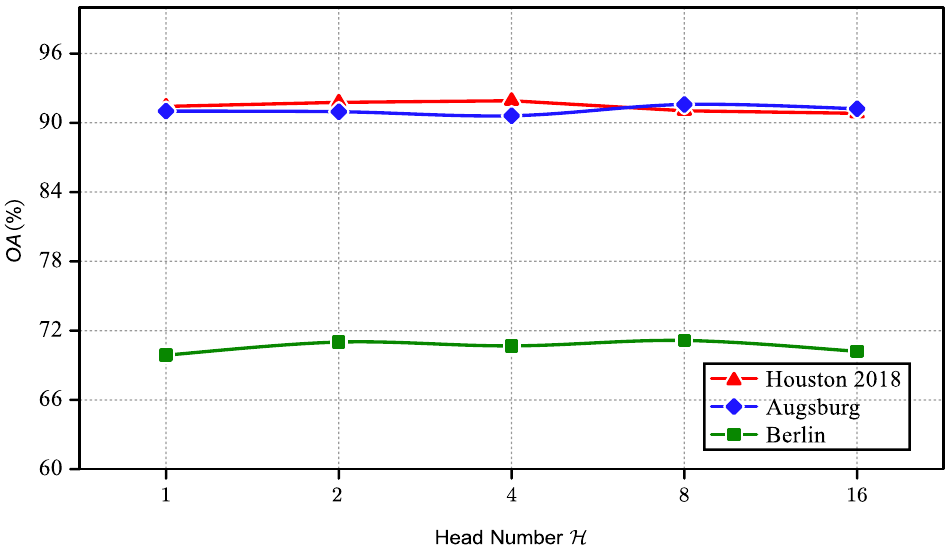}}
\caption{OA (\%) achieved by the proposed SS-MAE with different parameters for the Berlin, Augsburg, and Houston 2018 datasets. (a) Block number versus OA. (b) Token dimension versus OA. (c) Head number versus OA.}
\label{net_compare}
\end{figure*}
\subsubsection{Houston 2018 Dataset}
This dataset covers the University of Houston campus and the neighboring urban area. The HSI contains 48 bands in the wavelength range of 380–1050 nm. LiDAR data is a multispectral LiDAR image with three bands. The University of Houston released the dataset as part of the 2018 IEEE GRSS Data Fusion Contest. We use the training subset of the whole dataset in this paper. 

Table \ref{table_berlin}, Table \ref{table_augsburg}, and Table \ref{table_Houston2018} enumerate the number of samples for both training and testing on three datasets. Fig. \ref{dataset} displays the HSI data in false color, along with the SAR / LiDAR images and the ground truth. The evaluation metrics used in this paper include Overall Accuracy (OA), Average Accuracy (AA), and Kappa. Specifically, OA denotes the proportion of accurately classified samples throughout the entirety of the dataset, AA signifies the mean accuracy across all classes present within the dataset, and Kappa serves as a statistical measure of the concurrence between predicted and true labels.

\begin{figure}[htpb]
\centering
\includegraphics [width=0.45\textwidth]{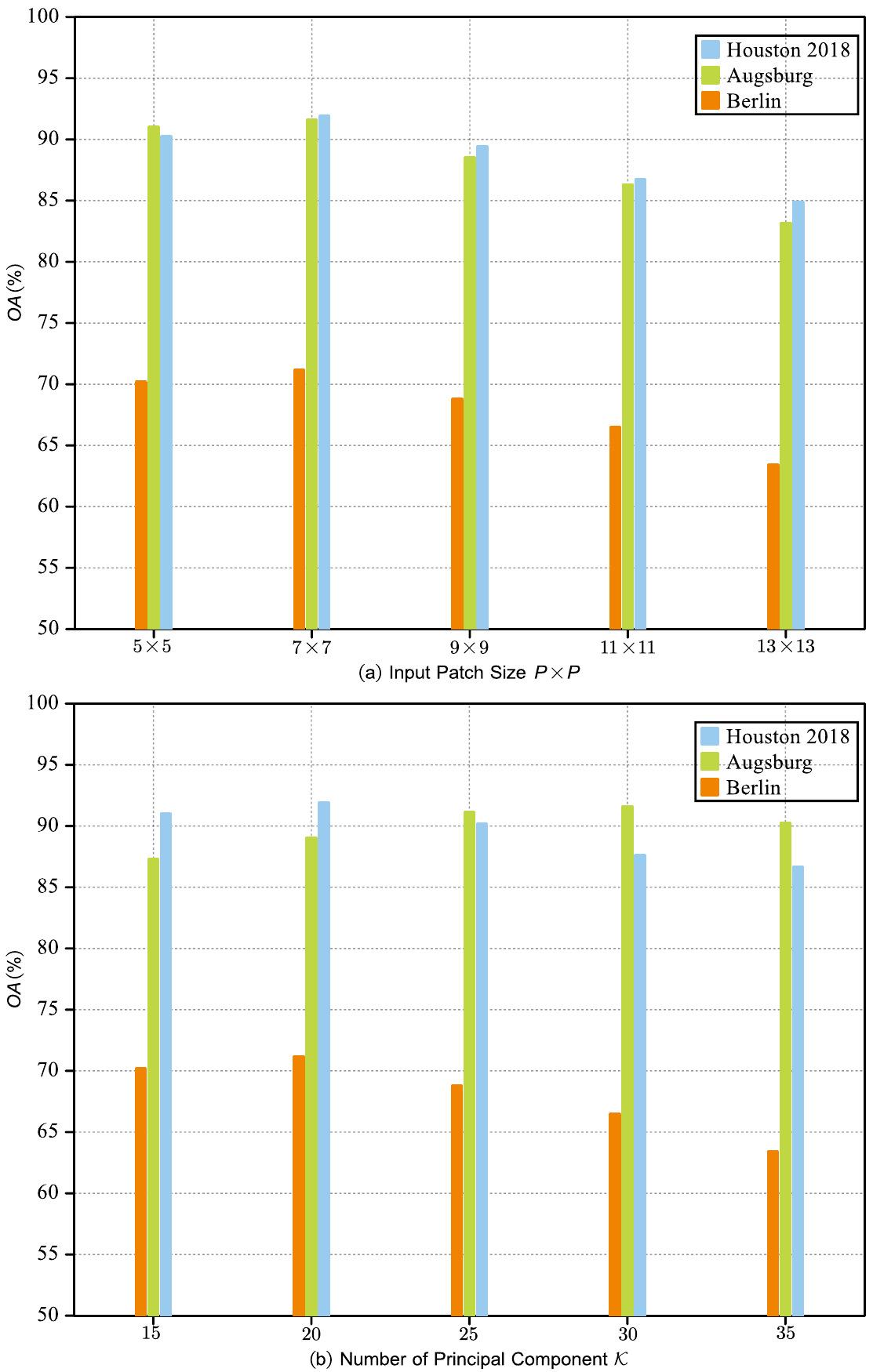}
\caption{OA (\%) achieved by the proposed SS-MAE with different parameters for the Berlin, Augsburg, and Houston 2018 datasets.}
\label{patch_pca}
\end{figure}
\subsection{Parameter Settings}

\begin{figure*}[htpb]
\centering
\includegraphics [width=0.99\textwidth]{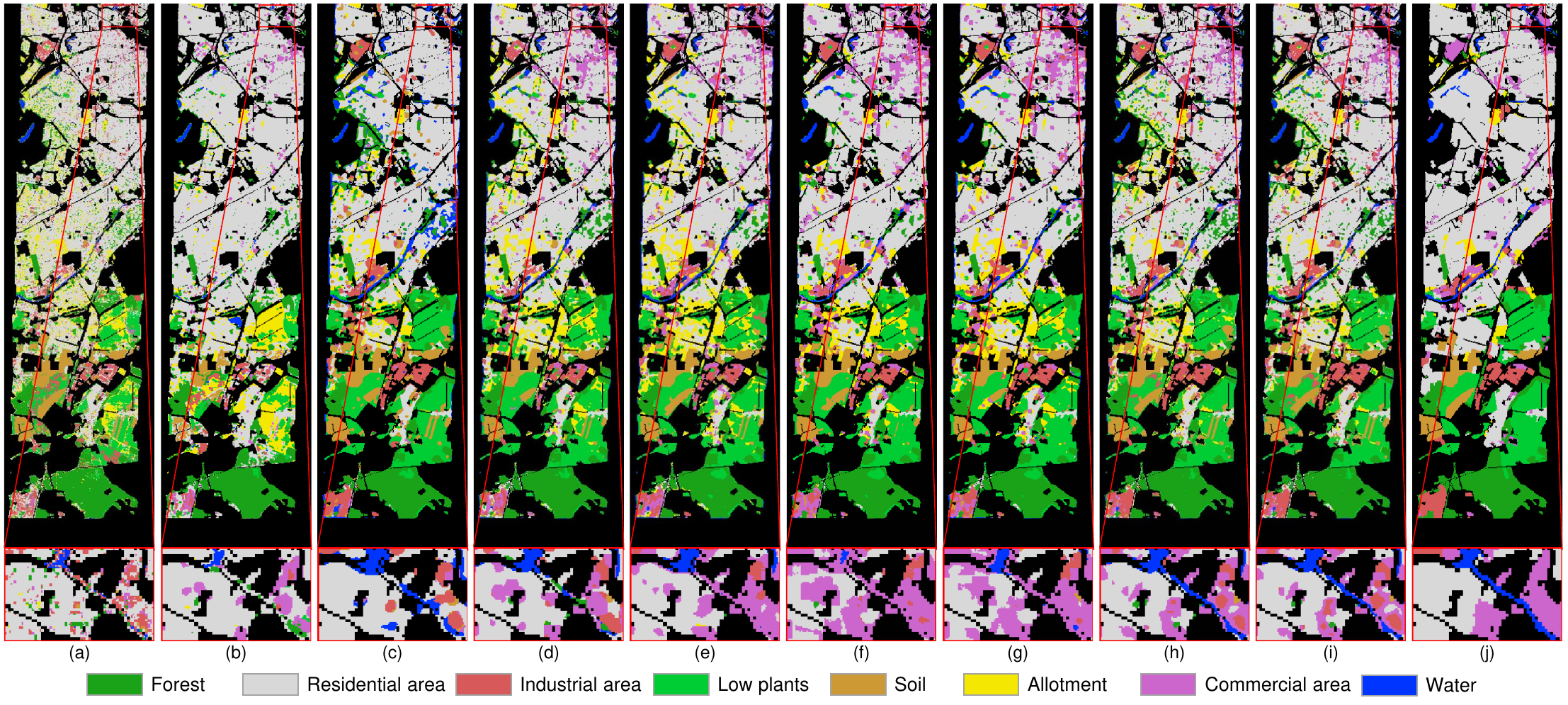}
\caption{Classification results of different methods for the Berlin dataset. (a) TBCNN. (b) FusAtNet. (c) $S^2$ENet. (d) DFINet. (e) AsyFFNet. (f) ExViT. (g) Fusion-HCT. (h) SS-MAE. (i) SS-MAE (Pre-trained). (j) Ground Truth.}
\label{berlin_fig_com}
\end{figure*}

\begin{figure*}[htpb]
\centering
\includegraphics [width=0.99\textwidth]{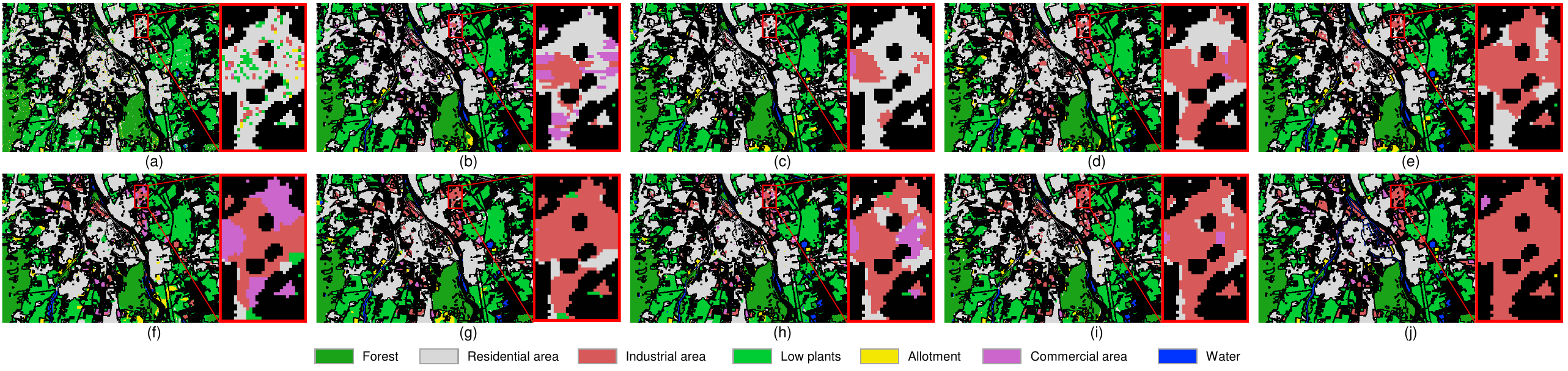}
\caption{Classification results of different methods for the Augsburg dataset. (a) TBCNN. (b) FusAtNet. (c) $S^2$ENet. (d) DFINet. (e) AsyFFNet. (f) ExViT. (g) Fusion-HCT. (h) SS-MAE. (i) SS-MAE (Pre-trained). (j) Ground Truth.}
\label{augsburg_fig_com}
\end{figure*}

\begin{figure*}[htpb]
\centering
\includegraphics [width=0.99\textwidth]{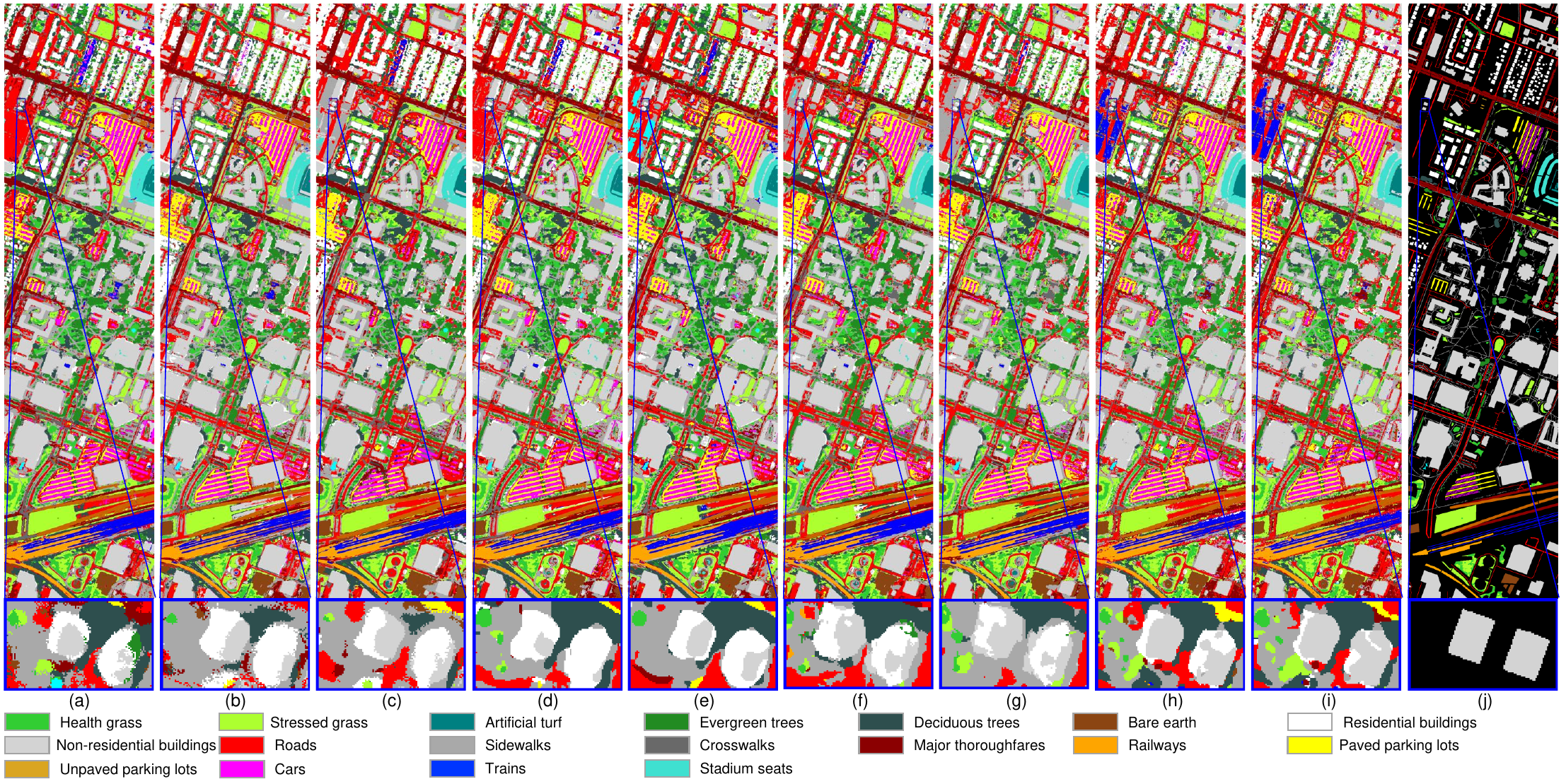}
\caption{Classification results of different methods for the Houston 2018 dataset. (a) TBCNN. (b) FusAtNet. (c) $S^2$ENet. (d) DFINet. (e) AsyFFNet. (f) ExViT. (g) Fusion-HCT. (h) SS-MAE. (i) SS-MAE (Pre-trained). (j) Ground Truth.}
\label{houston2018_fig_com}
\end{figure*}

This paper investigates the influence of parameter settings on classification performance for remote sensing image classification tasks and mask-reconstruction tasks. We fine-tune two categories of parameters: network structure (the number of blocks and heads in the Transformer encoder-decoder, and the size of feature dimension) and input data (input patch size, the number of principal components, the pre-training sample size, the spatial masking ratio, and the channel masking ratio).

\textbf{Parameters of the Network Structure.} To expedite experimentation, we maintain identical network structure parameters for the Transformer decoder and encoder throughout pre-training, and avoid pre-trained weights in preference to training from scratch in the experimental process for assessing the influence of network structure parameters.

\subsubsection{Block Number} Fig. \ref{net_compare}(a) shows the effect of different block number $\mathcal{B}$ values on the classification performance of on three datasets. When $\mathcal{B}=2$, our model achieved the highest OA values, which were 71.15\%, 91.59\%, and 91.91\%, respectively. It indicates that two blocks are sufficient to capture the essential features from the input data and enhance the representation power of SS-MAE.

\subsubsection{Token Dimension}
We test the optimal token dimension $d$ for each dataset. Fig. \ref{net_compare}(b) reveals the relationship between different $d$ values and the classification performance of the proposed SS-MAE for each dataset. It shows that when $d=256$, our model achieved the best OA values on all three datasets. It demonstrates that 256-dimensional feature for one token is optimal for the semantic information extraction from the input data.

\subsubsection{Heads Number}
We test the performance of the proposed SS-MAE with the head number $\mathcal{H}$ from 1 to 16. Fig. \ref{net_compare}(c) shows how different $\mathcal{H}$ values affect the classification performance on three datasets. The best $\mathcal{H}$ values for Berlin, Augsburg, and Houston 2018 datasets are 8, 8, and 4, respectively. Increasing the number of heads $\mathcal{H}$ does not significantly enhance performance and sometimes even degrades the classification accuracy.

\begin{table}[htpb] \centering \caption{Performance of the proposed model with different pre-training sample numbers}
\begin{tabular}{c|c|c|c} \toprule 
\multirow{2}{*}{Pre-training sample number} & \multicolumn{3}{c}{Overall Accuracy (\%)} \\ \cmidrule(lr){2-4}
& Berlin & Augsburg & Houston 2018\\ \midrule 
1000 & 71.32 & 92.11 & 91.93\\ 
10000 & 71.59 & 92.21 & 91.89\\ 
50000 & 71.91 & \textbf{92.27} & \textbf{92.43}\\ 
~100000~ & 72.36 & 92.26 & 92.25\\
~200000~ & \textbf{73.43} & - & 92.26\\ \bottomrule \end{tabular} \label{pretrain_sample} \end{table}

\begin{table}[htpb] \centering \caption{Performance of the proposed model with different masking ratio}
\begin{tabular}{c|c|c|c} \toprule 
\multirow{2}{*}{Masking Ratio} & \multicolumn{3}{c}{Overall Accuracy(\%)} \\ \cmidrule(lr){2-4}

& Berlin & Augsburg & Houston 2018\\ \midrule 
0.1 & 72.08 & 92.17 & 91.99\\ 
0.3 & \textbf{73.43} & 92.24 & 92.32\\ 
0.5 & 71.85 & 92.25 & \textbf{92.43}\\ 
0.7 & 72.08 & \textbf{92.27} & 92.29\\
0.9 & 72.36 & 92.20 & 92.15\\ \bottomrule \end{tabular} \label{mask_ratio} \end{table}

\setcounter{subsubsection}{0}

\textbf{Parameters of the Input Data.} In addition to the regular parameters of the input patch size, and the number of principal components, we pay special attention to three parameters against training data in pre-training stage: the pre-training sample size, the spatial masking ratio, and the channel masking ratio. We keep the spatial masking ratio and channel masking ratio consistent during pre-training.

\subsubsection{Input Patch Size}
We test five input patch sizes: $5 \times 5$, $7 \times 7$, $9 \times 9$, $11 \times 11$, and $13 \times 13$. The input patch size determines the spatial receptive fields 
captured by each token. As shown in Fig. \ref{patch_pca}(a), the optimal input patch size for all three datasets is $7 \times 7$. This indicates that a moderate amount of spatial information is sufficient for our model to learn effective representations, and larger patches increase the redundancy of information.

\definecolor{n1}{RGB}{255,0,0}
\definecolor{n2}{RGB}{0,0,255}
\definecolor{n3}{RGB}{0,255,0}

\begin{table*}[hptb]
\renewcommand\arraystretch{1.4}
\centering
\caption{Classification performance of different methods on the Berlin dataset.}
\begin{tabular}{c|cccccccccc}
\toprule  
	~~~Class~~~ & ~TBCNN~ & FusAtNet & ~$S^2$ENet~ & ~DFINet~ & AsyFFNet & ExViT & Fusion-HCT& SS-MAE$^*$ & SS-MAE\\\hline
	Forest & ~81.75~ & ~86.24~ & ~81.09~ & ~80.90~ & ~77.40~ & ~78.01~ & ~74.58~ & ~79.53~ & ~79.35~ \\ 
	Residential area & ~76.26~ & ~91.38~ & ~73.05~ & ~72.81~ & ~71.30~ & ~74.05~ & ~72.58~ & ~70.99~ & ~76.04~ \\ 
    Industrial area & ~39.67~ & ~19.76~ & ~62.61~ & ~38.89~ & ~38.98~ & ~39.48~ & ~48.92~ & ~63.06~ & ~68.51~ \\
    Low plants & ~49.78~ & ~20.00~ & ~82.82~ & ~78.09~ & ~79.30~ & ~84.15~ & ~83.04~ & ~80.19~ & ~79.17~ \\
    Soil & ~89.42~ & ~48.72~ & ~86.41~ & ~73.48~ & ~87.69~ & ~88.03~ & ~75.24~ & ~92.87~ & ~91.59~ \\
    Allotment & ~54.36~ & ~38.89~ & ~54.61~ & ~72.54~ & ~81.65~ & ~70.00~ & ~76.34~ & ~64.60~ & ~56.86~ \\
    Commercial area & ~4.65~ & ~18.47~ & ~2.56~ & ~22.80~ & ~36.52~ & ~38.18~ & ~29.66~ & ~26.37~ & ~19.75~ \\
    Water & ~41.93~ & ~29.61~ & ~75.96~ & ~68.15~ & ~75.96~ & 
 ~56.41~ & ~78.50~ & ~68.40~ & ~66.87~ \\\hline
    OA & ~67.60~ & ~70.91~ & ~71.08~ & ~70.33~ & ~70.82~ & ~72.63~ & ~71.18~ & ~71.15~ & ~73.43~ \\ 
    AA & ~54.72~ & ~44.13~ & ~64.88~ & ~63.45~ & ~68.59~ & ~66.04~ & ~67.36~ & ~68.24~ & ~67.26~ \\ 
    Kappa & ~50.96~ & ~51.07~ & ~58.02~ & ~56.90~ & ~58.53~ & ~60.48~ & ~58.66~ & ~59.00~ & ~60.91~ \\ 
\bottomrule
\end{tabular}
\label{berlin_compare_table}
\end{table*}

\subsubsection{Number of Principal Components} We use principal component analysis (PCA) to reduce the spectral redundancy of the input HSI. We test different number of principal components $\mathcal{K}$ from 15 to 35. Fig. \ref{patch_pca}(b) shows how $\mathcal{K}$ affects the classification performance on different datasets. It indicates that $\mathcal{K}=30$ is optimal for Berlin Dataset and Augsburg Dataset, while $\mathcal{K}=20$ is optimal for Houston 2018 Dataset. Therefore, we draw the conclusion that different datasets have different levels of spectral complexity and redundancy.

\subsubsection{Pre-Training Sample Number} For pre-training, we randomly sample a subset of pixels from each dataset. Table \ref{pretrain_sample} illustrates how increasing the sample number $\mathcal{A}$ improves the classification performance of our model on each dataset. We found that $\mathcal{A} = 50000$ is optimal for Augsburg Dataset and Houston 2018 Dataset, while $\mathcal{A} = 200000$ is optimal for Berlin Dataset. Overall, we observe that a larger sample set leads to better performance. Note that the classification accuracy tends to decrease slightly when $\mathcal{A}$ is greater than 50000 on Augsburg and Houston 2018 datasets. The optimal sample number for pre-training depends on the diversity of each dataset.

\subsubsection{The Spatial / Channel Masking Ratio} We use two types of masking strategies during pre-training: spatial masking (corrupting contiguous regions) and channel masking (corrupting entire spectral bands). We keep both masking ratios consistent during pre-training. We test different masking ratios from  0.1 to 0.9. Table \ref{mask_ratio} depicts the relationship between masking ratio and classification accuracy on three datasets. The Berlin, Augsburg, and Houston 2018 datasets require optimal masking ratios of 0.3, 0.7, and 0.5 respectively. It is evident that with a higher masking ratio, the model learns more robust feature representations via sparse data reconstruction. Nevertheless, if the ratio is too high, the model may fail to restore comprehensive information and impair the classification performance. 

\definecolor{n1}{RGB}{255,0,0}
\definecolor{n2}{RGB}{0,0,255}
\definecolor{n3}{RGB}{0,255,0}


\begin{table*}[t]
\renewcommand\arraystretch{1.4}
\centering
\caption{Classification performance of different methods on the Augsburg dataset.}
\begin{tabular}{c|cccccccccc}
\toprule
	~~~Class~~~ & ~TBCNN~ & FusAtNet & ~$S^2$ENet~ & ~DFINet~ & AsyFFNet & ExViT & Fusion-HCT& SS-MAE$^*$ & SS-MAE\\\hline
	Forest & ~90.88~ & ~93.78~ & ~98.10~ & ~97.38~ & ~96.01~ & ~90.04~ & ~97.32~ & ~96.82~ & ~96.48~ \\ 
	Residential area & ~93.89~ & ~97.58~ & ~99.08~ & ~98.37~ & ~97.96~ & ~95.44~ & ~98.28~ & ~98.14~ & ~97.77~ \\ 
    Industrial area & ~8.28~ & ~26.48~ & ~12.19~ & ~61.31~ & ~53.29~ & ~34.58~ & ~67.12~ & ~55.32~ & ~74.77~ \\
    Low plants & ~91.97~ & ~97.67~ & ~91.78~ & ~92.63~ & ~91.74~ & ~90.68~ & ~93.33~ & ~95.94~ & ~95.87~ \\
    Allotment & ~38.24~ & ~52.77~ & ~45.12~ & ~49.33~ & ~65.39~ & ~51.82~ & ~56.41~ & ~45.80~ & ~48.09~ \\
    Commercial area & ~1.40~ & ~24.66~ & ~1.22~ & ~3.54~ & ~4.58~ & ~28.63~ & ~8.42~ & ~14.29~ & ~9.45~ \\
    Water & ~10.82~ & ~47.51~ & ~24.09~ & ~26.61~ & ~49.64~ & 
 ~17.65~ & ~49.77~ & ~46.21~ & ~43.52~ \\\hline
    OA & ~84.53~ & ~90.62~ & ~88.22~ & ~90.66~ & ~90.15~ & 
 ~86.65~ & ~91.75~ & ~91.59~ & ~92.27~ \\ 
    AA & ~47.92~ & ~62.92~ & ~53.08~ & ~61.30~ & ~65.51~ & 
 ~58.40~ & ~67.24~ & ~64.64~ & ~66.56~ \\ 
    Kappa & ~77.13~ & ~86.33~ & ~82.61~ & ~86.47~ & ~85.74~ & 
 ~80.79~ & ~88.12~ & ~87.67~ & ~88.68~ \\ 
\bottomrule
\end{tabular}
\label{augsburg_compare_table}
\end{table*}

\begin{table*}[ht]
\renewcommand\arraystretch{1.4}
\centering
\caption{Classification performance of different methods on the Houston 2018 dataset.}
\scalebox{0.99}{
\begin{tabular}{c|cccccccccc}
\toprule

	~Class~ & TBCNN & FusAtNet & $S^2$ENet & DFINet & AsyFFNet & ExViT & Fusion-HCT& SS-MAE$^*$ & SS-MAE\\\hline
	Health grass & ~94.84~ & ~{96.28}~ & ~90.93~ & ~94.29~ & ~96.90~ & ~93.65~ & ~98.44~ & ~90.53~ & ~94.00~ \\ 
	Stressed grass & ~92.60~ & ~{93.45}~ & ~{94.76}~ & ~92.71~ & ~91.82~& ~95.44~ & ~91.05~ & ~95.56~& ~95.34~\\ 
    Artificial turf  & ~{100.00}~ & ~{100.00}~ & ~{100.00}~ & ~{100.00}~ & ~100.00~& ~100.00~ & ~100.00~ & ~100.00~& ~100.00~\\
    Evergreen trees & ~{98.80}~ & ~98.33~ & ~98.52~ & ~98.92~ & ~99.29~& ~98.52~ & ~97.39~ & ~98.23~ & ~98.55~\\ 
    Deciduous trees & ~97.12~ & ~{99.11}~ & ~{99.39}~ & ~97.77~ & ~99.52~& ~99.25~ & ~98.41~ & ~96.62~& ~98.15~\\ 
    Bare earth & ~99.61~ & ~{100.00}~ & ~{99.96}~ & ~100.00~ & ~99.98~& ~99.92~ & ~100.00~ & ~100.00~& ~99.84~\\ 
    Water & ~{100.00}~ & ~{100.00}~ & ~{100.00}~ & ~{100.00}~ & ~100.00~& ~100.00~ & ~100.00~ & ~100.00~& ~100.00~\\ 
    Residential buildings  & ~93.21~ & ~{97.90}~ & ~{97.16}~ & ~97.36~ & ~95.16~& ~96.88~ & ~96.80~ & ~95.89~& ~94.98~\\ 
    Non-residential buildings & ~91.30~ & ~93.41~ & ~{94.93}~ & ~93.29~ & ~95.02~& ~94.87~ & ~94.74~ & ~95.85~& ~96.27~\\ 
    Roads & ~61.06~ & ~{74.56}~ & ~{74.58}~ & ~75.73~ & ~72.72~& ~80.51~ & ~76.99~ & ~78.77~& ~82.97~\\ 
    Sidewalks & ~75.91~ & ~{82.94}~  & ~79.47~ & ~83.67~ & ~77.18~& ~80.27~ & ~83.14~ & ~78.29~& ~74.10~\\ 
    Crosswalks & ~{85.31}~ & ~84.65~ & ~{96.36}~ & ~94.23~ & ~93.14~& ~96.76~ & ~93.89~ & ~97.78~& ~89.57~\\ 
    Major thoroughfares & ~72.77~ & ~{86.90}~ & ~83.50~ & ~81.20~ & ~84.67~& ~82.11~ & ~86.29~ & ~80.20~& ~83.21~\\ 
    Highways & ~95.86~ & ~97.59~ & ~{98.57}~ & ~98.91~ & ~99.53~& ~84.10~ & ~93.75~ & ~98.28~& ~98.08~\\ 
    Railways & ~{99.78}~ & ~99.71~ & ~{99.84}~ & ~99.94~ & ~99.90~& ~99.81~ & ~99.86~ & ~100.00~& ~99.73~\\ 
    Paved parking lots & ~90.74~ & ~{98.08}~ & ~{97.41}~ & ~98.37~ & ~99.18~& ~99.08~ & ~98.62~ & ~95.36~& ~96.36~\\ 
    Unpaved parking lots & ~{100.00}~ & ~{100.00}~ & ~{100.00}~ & ~100.00~ & ~100.00~& ~100.00~ & ~100.00~ & ~100.00~& ~100.00~\\ 
    Cars & ~{98.47}~ & ~97.11~ & ~{99.17}~ & ~99.09~ & ~96.64~ & ~97.56~ & ~95.78~ & ~98.19~ & ~97.20~\\ 
    Trains & ~99.90~ & ~99.63~ & ~{100.00}~ & ~99.46~ & ~100.00~& ~99.90~ & ~100.00~ & ~99.46~& ~100.00~\\ 
    Stadium seats & ~99.92~ & ~99.93~ & ~{99.98}~ & ~99.98~ & ~99.89~& ~99.94~ & ~99.97~ & ~100.00~& ~100.00~\\\hline
    OA & ~86.95~ & ~{91.52}~ & ~{91.61}~ & ~91.02~ & ~91.24~& ~91.87~ & ~91.98~ & ~91.91~& ~92.43~\\ 
    AA & ~92.36~ & ~{94.98}~ & ~{95.22}~ & ~95.24~ & ~95.02~& ~94.93~ & ~95.26~ & ~94.94~& ~94.91~\\ 
    Kappa & ~83.39~ & ~{89.14}~ & ~{89.17}~ & ~88.46~ & ~88.71~& ~89.51~ & ~89.66~ & ~89.51~& ~90.19~\\ 
\bottomrule
\end{tabular}
}
\label{houston2018_compare_table}
\end{table*}

\begin{table}[t]
\renewcommand\arraystretch{1.4}
\centering
\caption{Quantitative Comparison of Model Complexity in Terms of the Number of Parameters and GFLOPs}
\begin{tabular}{c|cc}
\toprule
	Model & ~Params. (MB)~ & GFLOPs\\\hline
	TBCNN & ~0.1059~ & ~0.6567~\\ 
	FusAtNet & ~36.1518~ & ~305.9769~ \\ 
    $S^2$ENet & ~0.1629~ & ~1.7582~\\ 
    DFINet & ~0.7968~ & ~5.4959~\\ 
    AsyFFNet & ~1.0632~ & ~8.6556~\\ 
    ExViT & ~0.2270~ & ~2.9085~\\ 
    Fusion-HCT & ~0.4310~ & ~0.8057~\\ 
    SS-MAE & ~4.4000~ & ~15.0728~\\
\bottomrule
\end{tabular}
\label{complexity_compare}
\end{table}

\subsection{Performance Comparison}

We compare the proposed SS-MAE with existing state-of-the-art methods, including TBCNN \cite{xu2017multisource}, FusAtNet \cite{mohla2020fusatnet}, $S^2$ENet \cite{fang2021s2enet}, DFINet \cite{gao2021hyperspectral}, AsyFFNet \cite{li2022asymmetric}, ExViT \cite{yao2023extended}, and Fusion-HCT \cite{zhao2022joint}. TBCNN investigates the classification fusion of hyperspectral imagery and data from other sensors using a two-branch convolution neural network, which extracts spectral-spatial features from HSI and other data features from LiDAR or high-resolution visual images. FusAtNet proposes a multi-source classification framework by utilizing a self-attention mechanism for spectral features and a cross-attention approach for spatial features. $S^2$ENet proposes a spatial-spectral enhancement module for cross-modal information interaction, which effectively facilitates the interaction and understanding between hyperspectral and LiDAR data. DFINet uses a depth-wise cross attention module to extract complementary information from multisource feature pairs.  AsyFFNet is a multisource data classification method based on asymmetric feature fusion, which employs weight-share residual blocks for feature extraction and a feature calibration module for the spatial-wise multisource feature modeling. ExViT utilizes parallel branches of position-shared transformer extended with separable convolution modules to process multimodal image patches. A cross-modality attention module is employed to fuse tokenized embeddings. Fusion-HCT combines hierarchical CNN and transformer network to extract multisource data features, and then fuses both sets of features by cross-token attention fusion encoder.

\begin{table*}[h]
\centering
\caption{Influence of spectral attention, spatial attention, and IRB on classification results of SS-MAE.}
\setlength{\tabcolsep}{12pt}
\renewcommand{\arraystretch}{1.4}
\begin{tabular}{ccc|ccc} \hline
\toprule
\makecell{Spectral\\attention}&  \makecell{Spatial\\attention} & ~~~ IRB ~~~ & ~~~~ Berlin ~~~~ & Augsburg & Houston2018 \\
\midrule
\faCheck & {\color{gray}\faTimes} & {\color{gray}\faTimes} & 67.42 & 86.04 & 86.73\\ 
{\color{gray}\faTimes} & \faCheck & {\color{gray}\faTimes} & 70.73 & 89.92 & 90.81 \\ 
\faCheck & \faCheck & {\color{gray}\faTimes} & 72.35 & 90.91 & 91.73\\ 
\faCheck & \faCheck & \faCheck & 73.43 & 92.27 & 92.43\\   
\bottomrule \hline
\end{tabular}
\label{table_ablation_compare}
\end{table*}

\begin{table}[t]
\renewcommand\arraystretch{1.4}
\centering
\caption{Few-shot classification experiments of SS-MAE$^*$ and SS-MAE on Houston 2018 dataset.}
\begin{tabular}{c|cc}
\toprule
  Number of samples & ~SS-MAE$^*$~ & SS-MAE\\\hline
  1 & ~39.01~ & ~40.22~\\ 
  2 & ~39.91~ & ~42.37~ \\ 
  4 & ~54.17~ & ~55.57~\\  
  8 & ~61.39~ & ~62.25~\\ 
  16 & ~67.92~ & ~68.97~\\ 
\bottomrule
\end{tabular}
\label{few_shot}
\end{table}

\textit{1) Results on the Berlin Dataset.} The classification performance of different methods on the Berlin dataset is shown in Table \ref{berlin_compare_table}. SS-MAE$^*$ is proposed model without pre-training. Compared to other methods, SS-MAE$^*$ and SS-MAE achieve better classification performance in Industrial area class and Soil class. The OA improvements achieved by SS-MAE$^*$ are 3.55\%, 8.7\%, 0.24\%, 0.07\%, 0.82\%, and 0.33\% compared with TBCNN, FusAtNet, $S^2$ENet, DFINet, and AsyFFNet \cite{li2022asymmetric}, respectively. As can be observed from Table \ref{berlin_compare_table} that SS-MAE* exhibits a decrease of 1.48\% compared to ExViT and a marginal reduction of 0.03\% compared to Fusion-HCT. With the help of pretraining, SS-MAE not only surpasses ExViT and Fusion-HCT, but also attains the best classification performance. The improvement in performance can be attributed to the model's comprehensive integration of contextual and semantic information during the pretraining phase. With pretraining, SS-MAE is able to capture more universal features from extensive quantities of unlabeled data. These universal features enable the proposed SS-MAE to develop a more profound comprehension of the input multi-source data, thereby resulting in an enhancement of classification accuracy. The visualized classification results are shown in Fig. \ref{berlin_fig_com}. It is evident that the proposed SS-MAE effectively distinguishes the Industrial area and Residential area, as well as Allotment and Low plants.

\textit{2) Results on the Augsberg Dataset.} Table \ref{augsburg_compare_table} illustrates the classification performance of different methods on the Augsburg dataset. In comparison with other methods, SS-MAE$^*$ achieves the best OA and Kappa values except for slightly lower than Fusion-HCT. Furthermore, the classification performance of SS-MAE is further enhanced via pre-training, resulting in an increase of 0.68\% in OA value. The visualized classification results are shown in Fig. \ref{augsburg_fig_com}. It demonstrates that the proposed SS-MAE can correctly classify the Industrial area and Residential area, which is a difficult task for the other methods. 

\textit{3) Results on the Houston 2018 Dataset.} Table VII illustrates the classification performance of various models on the Houston 2018 dataset. Without pre-training, SS-MAE$^*$ achieves the second-best OA value. Pre-training the SS-MAE model further improves its performance, with an increase of 0.52\% in OA. The classification maps of different methods are shown in Fig. 9. Major Thoroughfares and Roads are challenging classes since their spectral responses are similar. It is evident that the proposed SS-MAE shows fewer misclassifications between both classes.

\textbf{4) Analysis of Computational Complexity.} 
Table \ref{complexity_compare} shows the computational complexity of different methods on the Houston 2018 dataset. As can be observed that the proposed method ranks second in terms of both parameters (Params) and giga floating-point operations (GFLOPs). It means that the computational complexity of the proposed SS-MAE surpasses most of the existing deep learning-based models. The reason is that SS-MAE is a typical unsupervised method, harnessing rich spectral and spatial representations via unsupervised learning. The larger model generalizes better for unsupervised learning. Therefore, the proposed SS-MAE uses a Transformer-like architecture with 4.4 MB parameters to obtain enhanced spectral and spatial representations. In practical scenarios, our model can concurrently compute multi-head attentions, thereby enhancing both training efficiency and inference speed.

\subsection{Ablation Study}
\textit{1) Effectiveness of Essential Components.} To evaluate the effectiveness of the essential components of our proposed SS-MAE, we conduct ablation studies on three datasets. The ablation studies aim to investigate the impact of spatial attention, spectral attention, and IRB. Table \ref{table_ablation_compare} exhibits the relationship between different components and OA values. When employing only spatial attention, the OA values on the Berlin, Augsburg, and Houston 2018 datasets are 67.42\% 86.04\%, and 86.73\%, respectively. However, when solely using spectral attention, the OA values on three datasets are 3.31\%, 3.88\%, and 4.08\% higher. It indicates that for HSI interpretation, the spectral information is critical and is an important factor for land cover classification. When simultaneously using spatial and spectral attention, the classification performance is further improved. It reflects the essential role of the two-branch structure of the proposed SS-MAE, and SS-MAE can simultaneously capture the rich spatial and spectral features. Furthermore, for the sake of balancing global and local features, we add IRB for local feature extraction. The IRB increases the classification performance by 1.08\%, 1.36\%, and 0.7\% on the three datasets, respectively. It demonstrates the effectiveness of supplementing local features with lightweight CNNs.

\textit{2) Few-Shot Classification.} To verify the effectiveness of pre-training, we compare the classification performance of SS-MAE$^*$ with SS-MAE using small number of training samples. Specifically, we train two models on the Houston2018 dataset, using randomly picked 2, 4, 8, and 16 samples, respectively. The results presented in Table \ref{few_shot} show that pretraining consistently boosts the classification performance of the proposed SS-MAE.

\section{Conclusion}

In this paper, we propose a novel framework, SS-MAE, which is a spatial-spectral masked auto-encoder for multi-source remote sensing image classification. In particular, it consists of spatial-wise branch and spectral-wise branch. The spatial branch masks random patches and reconstructs missing pixels, while the spectral-wise branch masks random spectral channels and reconstructs missing channels. Therefore, the proposed SS-MAE exploits both the spatial and spectral feature representations of the input hyperspectral data. Furthermore, to complement local features in the training stage, we add two lightweight CNNs for feature extraction. Both global and local features are used in feature modeling. Extensive experiments on the dataset demonstrate the superiority of the proposed SS-MAE over several state-of-the-art methods. In particular, we find that the proposed SS-MAE has excellent performance even without pre-training. The Transformer and CNN hybrid architecture benefit the cross-modal feature representation.

In the future, we plan to explore a more diverse cross-modal feature fusion strategy. As a pioneer study, we also plan to conduct more experiments on masked image modeling for multi-source remote sensing data classification.

\bibliography{source}

\begin{thebibliography}{10}
\providecommand{\url}[1]{#1}
\csname url@samestyle\endcsname
\providecommand{\newblock}{\relax}
\providecommand{\bibinfo}[2]{#2}
\providecommand{\BIBentrySTDinterwordspacing}{\spaceskip=0pt\relax}
\providecommand{\BIBentryALTinterwordstretchfactor}{4}
\providecommand{\BIBentryALTinterwordspacing}{\spaceskip=\fontdimen2\font plus
\BIBentryALTinterwordstretchfactor\fontdimen3\font minus
  \fontdimen4\font\relax}
\providecommand{\BIBforeignlanguage}[2]{{%
\expandafter\ifx\csname l@#1\endcsname\relax
\typeout{** WARNING: IEEEtran.bst: No hyphenation pattern has been}%
\typeout{** loaded for the language `#1'. Using the pattern for}%
\typeout{** the default language instead.}%
\else
\language=\csname l@#1\endcsname
\fi
#2}}
\providecommand{\BIBdecl}{\relax}
\BIBdecl

\bibitem{lihengkai22tgrs}
H.~Li, B.~Zhou, and F.~Xu, ``Variation analysis of spectral characteristics of
  reclamation vegetation in a rare earth mining area under environmental
  stress,'' \emph{IEEE Transactions on Geoscience and Remote Sensing}, vol.~60,
  pp. 1--12, 2022.

\bibitem{cui22grsl}
B.~Cui, X.~Li, J.~Wu, G.~Ren, and Y.~Lu, ``Tiny-scene embedding network for
  coastal wetland mapping using {Zhuhai-1} hyperspectral images,'' \emph{IEEE
  Geoscience and Remote Sensing Letters}, vol.~19, pp. 1--5, 2022.

\bibitem{anomalydetect23tgrs}
T.~Guo, L.~He, F.~Luo, X.~Gong, Y.~Li, and L.~Zhang, ``Anomaly detection of
  hyperspectral image with hierarchical antinoise mutual-incoherence- induced
  low-rank representation,'' \emph{IEEE Transactions on Geoscience and Remote
  Sensing}, vol.~61, pp. 1--13, 2023.

\bibitem{diffchange23tgrs}
F.~Luo, T.~Zhou, J.~Liu, T.~Guo, and X.~Gong, ``Multiscale diff-changed feature
  fusion network for hyperspectral image change detection,'' \emph{IEEE
  Transactions on Geoscience and Remote Sensing}, vol.~61, pp. 1--13, 2023.

\bibitem{10078892}
S.~Wang, D.~Hou, and H.~Xing, ``A self-supervised-driven open-set unsupervised
  domain adaptation method for optical remote sensing image scene
  classification and retrieval,'' \emph{IEEE Transactions on Geoscience and
  Remote Sensing}, vol.~61, pp. 1--15, 2023.

\bibitem{10070826}
Q.~Liu, J.~Peng, Y.~Ning, N.~Chen, W.~Sun, Q.~Du, and Y.~Zhou, ``Refined
  prototypical contrastive learning for few-shot hyperspectral image
  classification,'' \emph{IEEE Transactions on Geoscience and Remote Sensing},
  vol.~61, pp. 1--14, 2023.

\bibitem{10075072}
C.~Wang, L.~Zhang, W.~Wei, and Y.~Zhang, ``Dynamic super-pixel normalization
  for robust hyperspectral image classification,'' \emph{IEEE Transactions on
  Geoscience and Remote Sensing}, vol.~61, pp. 1--13, 2023.

\bibitem{hg23tgrs}
F.~Luo, L.~Zhang, B.~Du, and L.~Zhang, ``Dimensionality reduction with enhanced
  hybrid-graph discriminant learning for hyperspectral image classification,''
  \emph{IEEE Transactions on Geoscience and Remote Sensing}, vol.~58, no.~8,
  pp. 5336--5353, 2020.

\bibitem{9963700}
J.~Shi and H.~Jin, ``Riemannian nearest-regularized subspace classification for
  polarimetric {SAR} images,'' \emph{IEEE Geoscience and Remote Sensing
  Letters}, vol.~19, pp. 1--5, 2022.

\bibitem{9930788}
J.~Cheng, D.~Xiang, Q.~Yin, and F.~Zhang, ``A novel crop classification method
  based on the tensor-{GCN} for time-series {PolSAR} data,'' \emph{IEEE
  Transactions on Geoscience and Remote Sensing}, vol.~60, pp. 1--14, 2022.

\bibitem{9916296}
J.~Ni, C.~López-Martínez, Z.~Hu, and F.~Zhang, ``Multitemporal {SAR} and
  polarimetric {SAR} optimization and classification: Reinterpreting temporal
  coherence,'' \emph{IEEE Transactions on Geoscience and Remote Sensing},
  vol.~60, pp. 1--17, 2022.

\bibitem{dualview23tgrs}
T.~Guo, R.~Wang, F.~Luo, X.~Gong, L.~Zhang, and X.~Gao, ``Dual-view spectral
  and global spatial feature fusion network for hyperspectral image
  classification,'' \emph{IEEE Transactions on Geoscience and Remote Sensing},
  vol.~61, pp. 1--13, 2023.

\bibitem{hypergraph23tgrs}
Y.~Duan, F.~Luo, M.~Fu, Y.~Niu, and X.~Gong, ``Classification via
  structure-preserved hypergraph convolution network for hyperspectral image,''
  \emph{IEEE Transactions on Geoscience and Remote Sensing}, vol.~61, pp.
  1--13, 2023.

\bibitem{9940932}
J.~Geng, R.~Wang, and W.~Jiang, ``Polarimetric {SAR} image classification based
  on feature enhanced superpixel hypergraph neural network,'' \emph{IEEE
  Transactions on Geoscience and Remote Sensing}, vol.~60, pp. 1--12, 2022.

\bibitem{10075631}
K.~Wu, J.~Fan, P.~Ye, and M.~Zhu, ``Hyperspectral image classification using
  spectral–spatial token enhanced {Transformer} with hash-based positional
  embedding,'' \emph{IEEE Transactions on Geoscience and Remote Sensing},
  vol.~61, pp. 1--16, 2023.

\bibitem{multihopgraph23tgrs}
H.~Zhou, F.~Luo, H.~Zhuang, Z.~Weng, X.~Gong, and Z.~Lin, ``Attention multihop
  graph and multiscale convolutional fusion network for hyperspectral image
  classification,'' \emph{IEEE Transactions on Geoscience and Remote Sensing},
  vol.~61, pp. 1--14, 2023.

\bibitem{9928571}
G.~Liu, Y.~Li, Y.~Chen, R.~Shang, and L.~Jiao, ``{Pol-NAS}: A neural
  architecture search method with feature selection for {PolSAR} image
  classification,'' \emph{IEEE Journal of Selected Topics in Applied Earth
  Observations and Remote Sensing}, vol.~15, pp. 9339--9354, 2022.

\bibitem{6504845}
A.~Moreira, P.~Prats-Iraola, M.~Younis, G.~Krieger, I.~Hajnsek, and K.~P.
  Papathanassiou, ``A tutorial on synthetic aperture radar,'' \emph{IEEE
  Geoscience and Remote Sensing Magazine}, vol.~1, no.~1, pp. 6--43, 2013.

\bibitem{liao2014combining}
W.~Liao, R.~Bellens, A.~Pi{\v{z}}urica, S.~Gautama, and W.~Philips, ``Combining
  feature fusion and decision fusion for classification of hyperspectral and
  {LiDAR} data,'' in \emph{Proceedings of IEEE Geoscience and Remote Sensing
  Symposium (IGARSS)}, 2014, pp. 1241--1244.

\bibitem{bigdeli15}
B.~Bigdeli, F.~Samadzadegan, and P.~Reinartz, ``Fusion of hyperspectral and
  {LiDAR} data using decision template-based fuzzy multiple classifier
  system,'' \emph{International Journal of Applied Earth Observation and
  Geoinformation}, vol.~38, pp. 309--320, 2015.

\bibitem{jia21decision}
S.~Jia, Z.~Zhan, M.~Zhang, M.~Xu, Q.~Huang, J.~Zhou, and X.~Jia, ``Multiple
  feature-based superpixel-level decision fusion for hyperspectral and {LiDAR}
  data classification,'' \emph{IEEE Transactions on Geoscience and Remote
  Sensing}, vol.~59, no.~2, pp. 1437--1452, 2021.

\bibitem{chen2017deep}
Y.~Chen, C.~Li, P.~Ghamisi, X.~Jia, and Y.~Gu, ``Deep fusion of remote sensing
  data for accurate classification,'' \emph{IEEE Geoscience and Remote Sensing
  Letters}, vol.~14, no.~8, pp. 1253--1257, 2017.

\bibitem{xu2017multisource}
X.~Xu, W.~Li, Q.~Ran, Q.~Du, L.~Gao, and B.~Zhang, ``Multisource remote sensing
  data classification based on convolutional neural network,'' \emph{IEEE
  Transactions on Geoscience and Remote Sensing}, vol.~56, no.~2, pp. 937--949,
  2017.

\bibitem{li2020a3clnn}
H.-C. Li, W.-S. Hu, W.~Li, J.~Li, Q.~Du, and A.~Plaza, ``{A3CLNN}: Spatial,
  spectral and multiscale attention {ConvLSTM} neural network for multisource
  remote sensing data classification,'' \emph{IEEE Transactions on Neural
  Networks and Learning Systems}, vol.~33, no.~2, pp. 747--761, 2022.

\bibitem{jia23cl}
S.~Jia, X.~Zhou, S.~Jiang, and R.~He, ``Collaborative contrastive learning for
  hyperspectral and {LiDAR} classification,'' \emph{IEEE Transactions on
  Geoscience and Remote Sensing}, vol.~61, pp. 1--14, 2023.

\bibitem{wang23nncl}
M.~Wang, F.~Gao, J.~Dong, H.-C. Li, and Q.~Du, ``Nearest neighbor-based
  contrastive learning for hyperspectral and {LiDAR} data classification,''
  \emph{IEEE Transactions on Geoscience and Remote Sensing}, vol.~61, pp.
  1--16, 2023.

\bibitem{bao2021beit}
H.~Bao, L.~Dong, S.~Piao, and F.~Wei, ``{BEiT}: {BERT} pre-training of image
  {Transformers},'' in \emph{Proceedings of International Conference on
  Learning Representations (ICLR)}, 2022, pp. 1--13.

\bibitem{he2022masked}
K.~He, X.~Chen, S.~Xie, Y.~Li, P.~Doll{\'a}r, and R.~Girshick, ``Masked
  autoencoders are scalable vision learners,'' in \emph{Proceedings of the
  IEEE/CVF Conference on Computer Vision and Pattern Recognition}, 2022, pp.
  16\,000--16\,009.

\bibitem{Girdhar_2023_CVPR}
R.~Girdhar, A.~El-Nouby, M.~Singh, K.~V. Alwala, A.~Joulin, and I.~Misra,
  ``Omnimae: Single model masked pretraining on images and videos,'' in
  \emph{Proceedings of the IEEE/CVF Conference on Computer Vision and Pattern
  Recognition (CVPR)}, June 2023, pp. 10\,406--10\,417.

\bibitem{cong2022satmae}
Y.~Cong, S.~Khanna, C.~Meng, P.~Liu, E.~Rozi, Y.~He, M.~Burke, D.~Lobell, and
  S.~Ermon, ``{SatMAE}: Pre-training {Transformers} for temporal and
  multi-spectral satellite imagery,'' in \emph{Proceedings of Advances in
  Neural Information Processing Systems (NeurIPS)}, 2022, pp. 197--211.

\bibitem{ge2019hyperspectral}
C.~Ge, Q.~Du, W.~Li, Y.~Li, and W.~Sun, ``Hyperspectral and {LiDAR} data
  classification using kernel collaborative representation based residual
  fusion,'' \emph{IEEE Journal of Selected Topics in Applied Earth Observations
  and Remote Sensing}, vol.~12, no.~6, pp. 1963--1973, 2019.

\bibitem{gaoyh22tgrs}
Y.~Gao, W.~Li, M.~Zhang, J.~Wang, W.~Sun, R.~Tao, and Q.~Du, ``Hyperspectral
  and multispectral classification for coastal wetland using depthwise feature
  interaction network,'' \emph{IEEE Transactions on Geoscience and Remote
  Sensing}, vol.~60, pp. 1--15, 2022.

\bibitem{9698196}
J.~Wang, J.~Li, Y.~Shi, J.~Lai, and X.~Tan, ``Am³net: Adaptive
  mutual-learning-based multimodal data fusion network,'' \emph{IEEE
  Transactions on Circuits and Systems for Video Technology}, vol.~32, no.~8,
  pp. 5411--5426, 2022.

\bibitem{he2020momentum}
K.~He, H.~Fan, Y.~Wu, S.~Xie, and R.~Girshick, ``Momentum contrast for
  unsupervised visual representation learning,'' in \emph{Proceedings of the
  IEEE/CVF Conference on Computer Vision and Pattern Recognition (CVPR)}, 2020,
  pp. 9729--9738.

\bibitem{chen2020simple}
T.~Chen, S.~Kornblith, M.~Norouzi, and G.~Hinton, ``A simple framework for
  contrastive learning of visual representations,'' in \emph{Proceedings of
  International Conference on Machine Learning (ICML)}, 2020, pp. 1597--1607.

\bibitem{caron2021emerging}
M.~Caron, H.~Touvron, I.~Misra, H.~J{\'e}gou, J.~Mairal, P.~Bojanowski, and
  A.~Joulin, ``Emerging properties in self-supervised vision transformers,'' in
  \emph{Proceedings of the IEEE/CVF International Conference on Computer Vision
  (CVPR)}, 2021, pp. 9650--9660.

\bibitem{liu2022multi}
C.~Liu, H.~Sun, Y.~Xu, and G.~Kuang, ``Multi-source remote sensing pretraining
  based on contrastive self-supervised learning,'' \emph{Remote Sensing},
  vol.~14, no.~18, p. 4632, 2022.

\bibitem{baevski2022data2vec}
A.~Baevski, W.-N. Hsu, Q.~Xu, A.~Babu, J.~Gu, and M.~Auli, ``Data2vec: A
  general framework for self-supervised learning in speech, vision and
  language,'' in \emph{Proceedings of the International Conference on Machine
  Learning (ICML)}, 2022, pp. 1298--1312.

\bibitem{10187150}
Z.~Feng, L.~Song, S.~Yang, X.~Zhang, and L.~Jiao, ``Cross-modal contrastive
  learning for remote sensing image classification,'' \emph{IEEE Transactions
  on Geoscience and Remote Sensing}, pp. 1--1, 2023.

\bibitem{vaswani2017attention}
A.~Vaswani, N.~Shazeer, N.~Parmar, J.~Uszkoreit, L.~Jones, A.~N. Gomez,
  L.~Kaiser, and I.~Polosukhin, ``Attention is all you need,'' in
  \emph{Proceedings of Advances in Neural Information Processing Systems
  (NeurIPS)}, vol.~30, 2017, pp. 1--15.

\bibitem{dosovitskiy2020image}
A.~Dosovitskiy, L.~Beyer, A.~Kolesnikov, D.~Weissenborn, X.~Zhai,
  T.~Unterthiner, M.~Dehghani, M.~Minderer, G.~Heigold, S.~Gelly, J.~Uszkoreit,
  and N.~Houlsby, ``An image is worth 16x16 words: Transformers for image
  recognition at scale,'' in \emph{Proceedings of International Conference on
  Learning Representations (ICLR)}, 2021, pp. 1--12.

\bibitem{10124835}
D.~Ye, Z.~Ni, H.~Wang, J.~Zhang, S.~Wang, and S.~Kwong, ``{CSformer}: Bridging
  convolution and {Transformer} for compressive sensing,'' \emph{IEEE
  Transactions on Image Processing}, vol.~32, pp. 2827--2842, 2023.

\bibitem{10091778}
H.~Zhang, F.~Mao, M.~Xue, G.~Fang, Z.~Feng, J.~Song, and M.~Song, ``Knowledge
  amalgamation for object detection with {Transformers},'' \emph{IEEE
  Transactions on Image Processing}, vol.~32, pp. 2093--2106, 2023.

\bibitem{10091771}
Z.~Chang, Z.~Feng, S.~Yang, and Q.~Gao, ``{AFT}: Adaptive fusion {Transformer}
  for visible and infrared images,'' \emph{IEEE Transactions on Image
  Processing}, vol.~32, pp. 2077--2092, 2023.

\bibitem{9989404}
T.~You, C.~Wu, Y.~Bai, D.~Wang, H.~Ge, and Y.~Li, ``{HMF-Former}:
  Spatio-spectral {Transformer} for hyperspectral and multispectral image
  fusion,'' \emph{IEEE Geoscience and Remote Sensing Letters}, vol.~20, pp.
  1--5, 2023.

\bibitem{9874815}
Y.~Peng, Y.~Zhang, B.~Tu, Q.~Li, and W.~Li, ``Spatial–spectral {Transformer}
  with cross-attention for hyperspectral image classification,'' \emph{IEEE
  Transactions on Geoscience and Remote Sensing}, vol.~60, pp. 1--15, 2022.

\bibitem{10138021}
X.~Tang, Y.~Wang, J.~Ma, X.~Zhang, F.~Liu, and L.~Jiao, ``Interacting-enhancing
  feature transformer for cross-modal remote-sensing image and text
  retrieval,'' \emph{IEEE Transactions on Geoscience and Remote Sensing},
  vol.~61, pp. 1--15, 2023.

\bibitem{10145469}
Y.~Zhang, S.~Xu, D.~Hong, H.~Gao, C.~Zhang, M.~Bi, and C.~Li, ``Multimodal
  transformer network for hyperspectral and lidar classification,'' \emph{IEEE
  Transactions on Geoscience and Remote Sensing}, vol.~61, pp. 1--17, 2023.

\bibitem{10153685}
S.~K. Roy, A.~Deria, D.~Hong, B.~Rasti, A.~Plaza, and J.~Chanussot,
  ``Multimodal fusion transformer for remote sensing image classification,''
  \emph{IEEE Transactions on Geoscience and Remote Sensing}, vol.~61, pp.
  1--20, 2023.

\bibitem{park2022vision}
N.~Park and S.~Kim, ``How do vision transformers work?'' in \emph{Proceedings
  of International Conference on Learning Representations (ICLR)}, 2022, pp.
  1--14.

\bibitem{hong2021multimodal}
D.~Hong, J.~Hu, J.~Yao, J.~Chanussot, and X.~X. Zhu, ``Multimodal remote
  sensing benchmark datasets for land cover classification with a shared and
  specific feature learning model,'' \emph{ISPRS Journal of Photogrammetry and
  Remote Sensing}, vol. 178, pp. 68--80, 2021.

\bibitem{okujeni2016berlin}
A.~Okujeni, S.~van~der Linden, and P.~Hostert, ``Berlin-urban-gradient dataset
  2009-an enmap preparatory flight campaign,'' \emph{GFZ Data Services}, 2016.

\bibitem{mohla2020fusatnet}
S.~Mohla, S.~Pande, B.~Banerjee, and S.~Chaudhuri, ``Fusatnet: Dual attention
  based spectrospatial multimodal fusion network for hyperspectral and lidar
  classification,'' in \emph{Proceedings of the IEEE/CVF Conference on Computer
  Vision and Pattern Recognition Workshops}, 2020, pp. 92--93.

\bibitem{fang2021s2enet}
S.~Fang, K.~Li, and Z.~Li, ``$s^2$enet: Spatial-spectral cross-modal
  enhancement network for classification of hyperspectral and lidar data,''
  \emph{IEEE Geoscience and Remote Sensing Letters}, 2021.

\bibitem{gao2021hyperspectral}
Y.~Gao, W.~Li, M.~Zhang, J.~Wang, W.~Sun, R.~Tao, and Q.~Du, ``Hyperspectral
  and multispectral classification for coastal wetland using depthwise feature
  interaction network,'' \emph{IEEE Transactions on Geoscience and Remote
  Sensing}, 2021.

\bibitem{li2022asymmetric}
W.~Li, Y.~Gao, M.~Zhang, R.~Tao, and Q.~Du, ``Asymmetric feature fusion network
  for hyperspectral and sar image classification,'' \emph{IEEE Transactions on
  Neural Networks and Learning Systems}, 2022.

\bibitem{yao2023extended}
J.~Yao, B.~Zhang, C.~Li, D.~Hong, and J.~Chanussot, ``Extended vision
  transformer (exvit) for land use and land cover classification: A multimodal
  deep learning framework,'' \emph{IEEE Transactions on Geoscience and Remote
  Sensing}, 2023.

\bibitem{zhao2022joint}
G.~Zhao, Q.~Ye, L.~Sun, Z.~Wu, C.~Pan, and B.~Jeon, ``Joint classification of
  hyperspectral and lidar data using a hierarchical cnn and transformer,''
  \emph{IEEE Transactions on Geoscience and Remote Sensing}, vol.~61, pp.
  1--16, 2022.

\end{thebibliography}
\bibliographystyle{IEEEtran}

\begin{IEEEbiography}[{\includegraphics[width=1in,height=1.25in,clip,keepaspectratio]{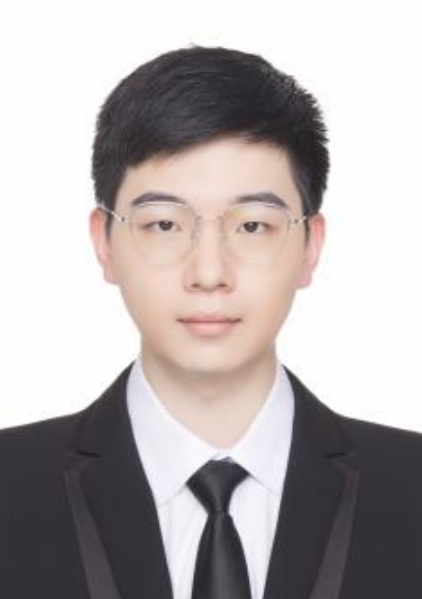}}]{Junyan Lin}
received the B.Sc. degree in computer science from Zhejiang Gongshang University, Hangzhou, China, in 2022. He is currently pursuing the M.Sc. degree in computer science and applied remote sensing with the School of Computer Science, Ocean University of China, Qingdao, China.

His current research interests include computer vision and remote sensing image processing.

\end{IEEEbiography}

\begin{IEEEbiography}[{\includegraphics[width=1in,height=1.25in,clip,keepaspectratio]{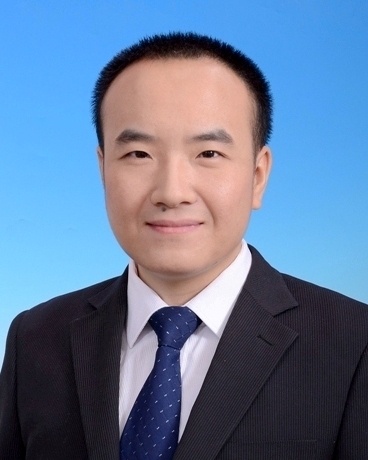}}]{Feng Gao} (Member, IEEE)
received the B.Sc degree in software engineering from Chongqing University, Chongqing, China, in 2008, and the Ph.D. degree in computer science and technology from Beihang University, Beijing, China, in 2015.

He is currently an Associate Professor with the School of Information Science and Engineering, Ocean University of China. His research interests include remote sensing image analysis, pattern recognition and machine learning.

\end{IEEEbiography}

\begin{IEEEbiography}[{\includegraphics[width=1in,height=1.25in,clip,keepaspectratio]{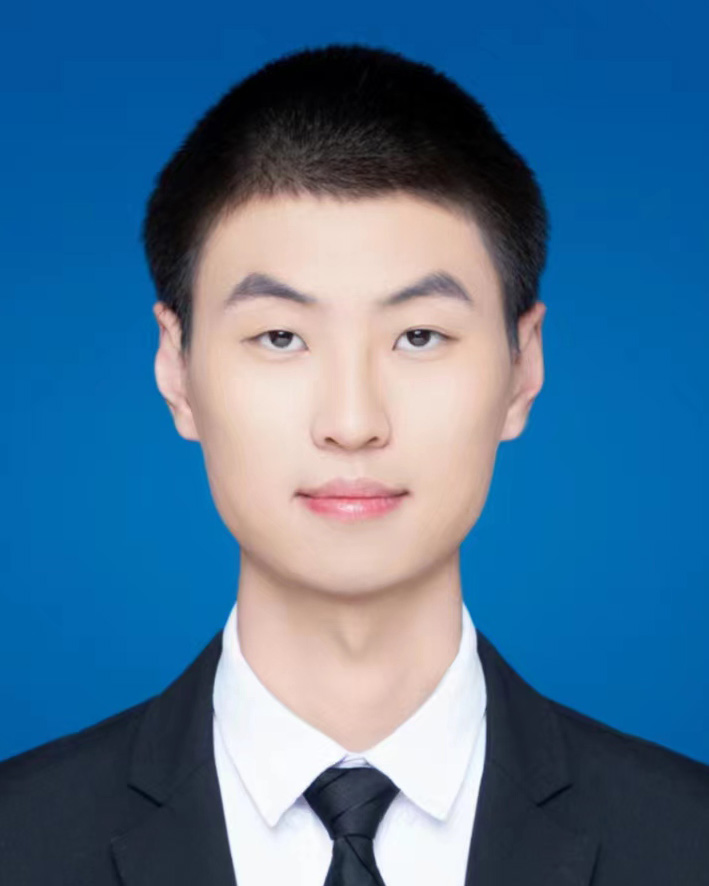}}]{Xiaochen Shi}
received the B.Sc. degree in computer science from Ocean University of China, Qingdao, China, in 2021. He is currently pursuing the M.Sc. degree in computer science and applied remote sensing with the School of Computer Science, Ocean University of China, Qingdao, China.

His current research interests include computer vision and remote sensing image processing.

\end{IEEEbiography}

\begin{IEEEbiography}[{\includegraphics[width=1in,height=1.25in,clip,keepaspectratio]{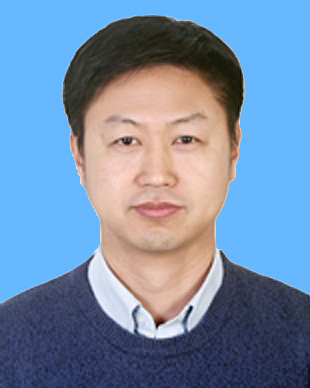}}]{Junyu Dong}
 (Member, IEEE) received the B.Sc. and M.Sc. degrees from the Department of Applied Mathematics, Ocean University of China, Qingdao, China, in 1993 and 1999, respectively, and the Ph.D. degree in image processing from the Department of Computer Science, Heriot-Watt University, Edinburgh, United Kingdom, in 2003.

He is currently a Professor and Dean with the School of Computer Science and Technology, Ocean University of China. His research interests include visual information analysis and understanding, machine learning and underwater image processing.
\end{IEEEbiography}

\begin{IEEEbiography}[{\includegraphics[width=1in,height=1.25in,clip,keepaspectratio]{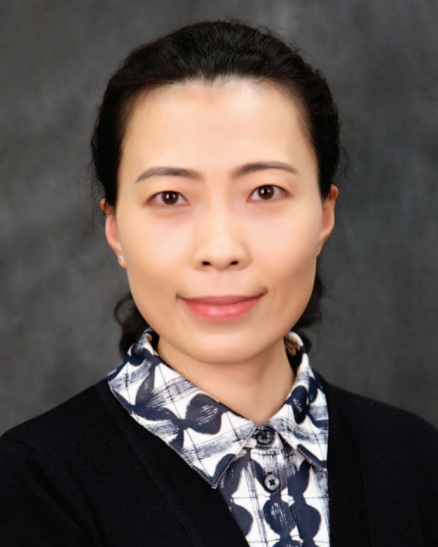}}]{Qian Du}
(Fellow, IEEE) received the Ph.D. degree in electrical engineering from the University of Maryland at Baltimore, Baltimore, MD, USA, in 2000.

She is currently the Bobby Shackouls Professor with the Department of Electrical and Computer Engineering, Mississippi State University, Starkville, MS, USA. Her research interests include hyperspectral remote sensing image analysis and applications, and machine learning. Dr. Du was the recipient of the 2010 Best Reviewer Award from the IEEE Geoscience and Remote Sensing Society (GRSS). She was a Co-Chair for the Data Fusion Technical Committee of the IEEE GRSS from 2009 to 2013, the Chair for the Remote Sensing and Mapping Technical Committee of International Association for Pattern Recognition from 2010 to 2014, and the General Chair for the Fourth IEEE GRSS Workshop on Hyperspectral Image and Signal Processing: Evolution in Remote Sensing held at Shanghai, China, in 2012. She was an Associate Editor for the \textsc{Pattern Recognition}, and \textsc{IEEE Transactions on Geoscience and Remote Sensing}. From 2016 to 2020, she was the Editor-in-Chief of the \textsc{IEEE Journal of Selected Topics in Applied Earth Observation and Remote Sensing}. She is currently a member of the IEEE Periodicals Review and Advisory Committee and SPIE Publications Committee. She is a Fellow of SPIE-International Society for Optics and Photonics (SPIE).

\end{IEEEbiography}

\end{document}